\documentclass[journal]{IEEEtran}
\usepackage{amssymb}
\usepackage{pifont}
\newcommand{\cmark}{\ding{51}}%
\newcommand{\xmark}{\ding{55}}%
\usepackage[abs]{overpic}
\usepackage{xcolor,varwidth}

\usepackage[noadjust]{cite}

\usepackage[hyphens]{url}
\usepackage{hyperref}

\ifCLASSINFOpdf

\else

\fi

\begin{document}
\title{Challenges of Driver Drowsiness Prediction: \\ The Remaining Steps to Implementation}

\author{Emma~Perkins,%
        ~Chiranjibi~Sitaula,%
        ~Michael~Burke,~\IEEEmembership{Member,~IEEE}
        ~and~Faezeh~Marzbanrad,~\IEEEmembership{Senior Member,~IEEE}%
\thanks{E. Perkins, C. Sitaula, M. Burke and F. Marzbanrad are with the Department
of Electrical and Computer Systems Engineering, Monash University, Clayton, 3800, VIC, Australia. E. Perkins' research is supported by an Australian Government Research Training Program (RTP) Scholarship.
}%
}

\maketitle

\begin{abstract}
Driver drowsiness has caused a large number of serious injuries and deaths on public roads and incurred billions of taxpayer dollars in costs. Hence, monitoring of drowsiness is critical to reduce this burden on society. This paper surveys the broad range of solutions proposed to address the challenges of driver drowsiness, and identifies the key steps required for successful implementation. Although some commercial products already exist, with vehicle-based methods most commonly implemented by automotive manufacturers, these systems may not have the level of accuracy required to properly predict and monitor drowsiness. State-of-the-art models use physiological, behavioural and vehicle-based methods to detect drowsiness, with hybrid methods emerging as a superior approach. Current setbacks to implementing these methods include late detection, intrusiveness and subject diversity. In particular, physiological monitoring methods such as Electroencephalography (EEG) are intrusive to drivers; while behavioural monitoring is least robust, affected by external factors such as lighting, as well as being subject to privacy concerns. Drowsiness detection models are often developed and validated based on subjective measures, with the Karolinska Sleepiness Scale being the most popular. Subjective and incoherent labelling of drowsiness, lack of on road data and inconsistent protocols for data collection are among other challenges to be addressed to progress drowsiness detection for reliable on-road use. 
\end{abstract}

\begin{IEEEkeywords}
drowsiness detection, fatigue, driver physiological monitoring, sensor applications, vehicle-based monitoring, behavioural monitoring, hybrid models, driver safety
\end{IEEEkeywords}

\IEEEpeerreviewmaketitle

\section{Introduction}

\IEEEPARstart{G}{lobally}, it is estimated that 1.3 million people die per year as a result of car crashes worldwide, where the crashes cost the country approximately 3\% of their GDP \cite{RN290}.

Fatigue has been labelled as part of the “fatal five” for driving safety risks, alongside speeding, drugs/alcohol, failure to wear a seat belt and driver distraction. To date, all but driving fatigue has been tackled by a legal requirement (e.g. speed limits/cameras, alcohol limit, seat belt regulations and phone usage restrictions whilst driving). Research has also shown and vigorously advertised that the effects of drowsiness when driving is similar to that of driving above the legal alcohol limit \cite{RN5, RN6}. Hence, the monitoring of fatigued driving is required to make driving safer and has emerged as a key priority to reduce road related fatalities and serious injuries.

In Australia, driving whilst fatigued has contributed to 20-30\% of road related severe injuries and fatalities, similar to that of drink driving and speeding \cite{RN1,RN2}. Australia has seen an average of 1,209 fatalities on road within the last 5 years (2016-2020) \cite{RN3}, with the Northern Territory producing the largest average fatality rate per 100,000 people. The percentage of serious injuries was shown to be higher for fatigue related instances, with road related accidents estimated to cost \$3-4 billion AUD to the economy in Victoria alone, per year \cite{RN4}. In 2015, it was reported that USA had an estimated 5000 people die in crashes influenced by drowsiness, although this is difficult to measure accurately \cite{RN135}. This was estimated to cost the economy \$109 billion USD \cite{RN135}. Based on reports from Australia, England and many European countries, approximately 10 to 30\% of crashes are caused by drowsiness \cite{RN289}. 

It is clear that driver fatigue remains a pressing concern for road safety. However, despite over a decade of research into detection and prediction measures, this problem has yet to be adequately addressed. This paper seeks to take stock of the progress made to date, and identifies key challenges holding back the implementation of driver drowsiness prediction technologies.
 
In preparing this paper, 126 studies were surveyed, including 43 Physiological-based studies, 40 Behavioural-based studies, 15 Vehicle-based studies, and 28 Hybrid studies. Of these studies, 65 were published after 2018 and considered to be recent. These studies were found by first searching for keywords covering the four topics listed in Table \ref{tab:keywords}. The selection was then expanded to include a greater number of recent studies, with the search restricted to 2018 and later. The search was conducted across Google Scholar, IEEEXplore, and ScienceDirect. 

\begin{table}[!t]
\renewcommand{\arraystretch}{1.3}
\caption{Search Terms for Drowsiness Studies}
\centering
\begin{tabular}{|c|c|}
\hline
\textbf{Method} & \textbf{Keywords} \\
\hline
All Studies & Driver, Fatigue, Drowsiness, Drowsy \\
\hline
Physiological & Physiological, PPG, ECG, EOG, \\ & EEG, EMG, Temperature, Respiration \\
\hline
Behavioural &  Behavioural, Camera, Yawning\\
\hline
Vehicle-Based  &  Vehicle, Lane departure, Steering Wheel \\
\hline
Hybrid & Hybrid, Physiological + Behavioural, \\
& Behavioural + Vehicle, Physiological + Vehicle \\
\hline
\end{tabular}
\label{tab:keywords}
\end{table}

\begin{table*}[!t]
\renewcommand{\arraystretch}{1.3}
\caption{Focus areas of recent review papers}
\centering
\begin{tabular}{ccccccccc}
\hline
\textbf{Paper} & \textbf{Features} & \textbf{Classification} & \textbf{Simulation vs} & \textbf{Labelling of} & \textbf{Prediction vs} & \textbf{Subject} & \textbf{Accuracy}\\
  & \textbf{/Modalities} & \textbf{Methods} & \textbf{on-road} & \textbf{drowsiness} & \textbf{detection} & \textbf{diversity} & \textbf{discrepancies}\\
\hline
Sahayadhas et al.,  & \cmark & \cmark & \cmark & \xmark & \xmark & \xmark & \xmark \\
2012 \cite{RN10} & & & & & & & & \\
\hline
Kaplan et al., & \cmark & \cmark & \cmark & \xmark & \xmark & \xmark & \xmark  \\
2015 \cite{RN279} & & & & & & & & \\
\hline
Doudou et al.,  & \cmark & \cmark & \xmark & \cmark & \xmark & \xmark & \xmark  \\
2019 \cite{RN36} & & & & & & & & \\
\hline
Ramzan et al.,  & \cmark & \cmark & \xmark & \xmark & \xmark & \xmark & \xmark  \\
2019 \cite{RN11} & & & & & & & & \\
\hline
Hu and Lodewijks,  & \cmark & \xmark & \cmark & \xmark & \xmark & \xmark & \xmark \\
2020 \cite{RN7} & & & & & & & & \\
\hline
Němcová et al.,  & \cmark & \cmark & \cmark & \cmark & \xmark & \xmark & \xmark \\
2021 \cite{RN29} & & & & & & & & \\
\hline
Ours & \xmark & \xmark & \cmark & \cmark & \cmark & \cmark & \cmark \\
 & & & & & & & & \\
\hline
\end{tabular}
\label{tab:reviews}
\end{table*}

Existing review papers have thoroughly covered technical aspects such as signal acquisition, feature extraction and detection of driver drowsiness, in three primary categories: physiological, behavioural and vehicle-based monitoring. Sahayadhas et al. \cite{RN10} outlines existing simulator types, causes of crashes, inducement of drowsiness, simulator versus on-road data collection and physiological intrusiveness. Further challenges regarding protocol are mentioned, where the environment for acquisition and drowsiness inducement are discussed; however, other issues including drowsiness labelling, early detection and subject diversity are not present in the review.

Kaplan et al. \cite{RN279} highlights ways to detect drowsiness, commercial products available and simulation data collection. Kaplan et al. also expands the review to driver distraction; another cause of deaths on our roads. Within the review, the physiological methods mention Electrooculography (EOG) and Electrocardiography (ECG); however, only EEG is discussed in depth as they label this as the most "promising and feasible method". Furthermore, limitations of described studies are mentioned, but not protocols including diversity between subjects and the number of subjects. 

DouDou et al. \cite{RN36} thoroughly examines acquisition systems, fatigue inducement and commercial products; however this differs from the focus of our work, as shown in Table \ref{tab:reviews}. Moreover, our paper includes a number of more recent works in the literature. DouDou et al. also looked at issues in the literature, including intrusiveness of acquisition, noise in data collection, ease of use and accuracy of different methods. They discuss the variations in fatigue measuring, circadian rhythm and the effect on models as well as the ability to track drowsiness in real-time. However, they do not discuss simulations versus on-road data, predicting drowsiness rather than detecting, and the issues with collected data including subject diversity, number of subjects and accuracy reporting differences; whereas our review address further challenges, with a focus on more recent studies. 

Ramzan et al. \cite{RN11} provides a very methodological, systematic review of the literature and accompanies fatigue features with examples from the literature. Classification methods are also expanded upon in the Ramzan et al. review; however, challenges are not a focus of this review, but rather identifying current studies on drowsiness and what methods are being used. 

Hu and Lodewijks \cite{RN7} provide an in-depth understanding of fatigue and sleepiness differentiation and how measures can effectively distinguish the two. Hu and Lodewijks also cover the downfalls of intrusive acquisition, the differences in results between simulator and on-road studies, and inconsistency of protocols. However, less intrusive acquisition systems (such as Photoplethysmography (PPG)) and hybrid studies are not discussed. 

Finally, Němcová et al. \cite{RN29} covers specific features and their link to drowsiness, mass produced drowsiness systems, privacy concerns, cost implications, and the annotation variances in drowsiness labels. They include a number of current challenges in driver drowsiness detection; however, early detection, specific experimental protocols (including the number of subjects and diversity of subjects), nor the exclusion of subjects with sleep disorders are discussed. 

Existing reviews demonstrate that the acquisition, feature selection and classification aspects of drowsiness monitoring have been well studied. In contrast, this paper focuses on barriers to the implementation of hybrid models, along with existing challenges. In particular, we focus on prediction vs detection, subject diversity, the number of subjects, accuracy reporting and future possibilities for driver drowsiness, which have not been adequately addressed to date. The current challenges in drowsiness detection covered by review papers is summarised in Table \ref{tab:reviews}. A key challenge identified in this work lies in the distinction between driver drowsiness prediction (prediction of impending drowsiness) and detection (recognising drowsiness at it occurs). The latter is most commonly addressed by current technologies, while the former is often neglected, although of significantly more value from a road safety perspective.

This paper is structured as follows. Initially, we provide a background of the effects of drowsiness and how it is monitored. Second, we define our evaluation metrics to review different drowsiness systems. We then describe existing commercial products for driver drowsiness detection, followed by methods under active research and development to provide an overview of recent literature. The active research methods introduce drowsiness labelling and monitoring categories for drowsiness detection. Existing challenges are then outlined, with explanations of how these challenges affect models and restrict these from becoming implementable. Future work is also outlined, providing a roadmap to more accurate and effective drowsiness technologies and enhanced road safety.

\begin{table*}[!t]
\renewcommand{\arraystretch}{1.3}
\caption{Performance of existing drowsiness detection methods}
\centering
\begin{tabular}{|c|c|c|c|c|c|c|c|c|c|}
\hline
\textbf{Method} & \textbf{Accuracy} & \textbf{Physical} & \textbf{Psychological} & \textbf{Privacy} & \textbf{Data} & \textbf{Cost} & \textbf{Susceptibility} & \textbf{Vulnerability to} & \textbf{Adaptability}\\
 & & \textbf{Intrusiveness} & \textbf{Intrusiveness} & \textbf{Concern} & \textbf{Size} & & \textbf{to noise} & \textbf{Subject Diversity} & \\
\hline
\textbf{Physiological}  & 1 & 3$^*$ & 1 & 1 & 2 & 3 & 3 & 2 & 1\\
\hline
\textbf{Behavioural} & 2 & 1 & 3 & 3 & 3 & 2 & 1 & 3 & 3 \\
\hline
\textbf{Vehicle-Based} & 3 & 1 & 2 & 2 & 1 & 1 & 2 & 1 &  2\\
\hline
\end{tabular}
\label{tab:ranking}
\flushleft\footnotesize{ ~\\$*$ This can vary depending on the acquisition system}
\end{table*}

\section{Background}
Drowsiness is the transitional phase between wake and sleep, where a person can be mentally and physically affected \cite{RN22}. Drowsiness has been shown to depend on the time of day (circadian rhythm) and lack or interruption of sleep \cite{RN22, RN23}. Furthermore, increased driving times, warm rooms and monotonous activities have been found to unmask sleepiness rather than cause it \cite{RN24}. Causes of drowsiness can include work shift schedules (where a decrease in performance has been seen in night shift workers \cite{RN23}), lifestyle choices, mental health problems, stress and medication or illness \cite{RN25, RN291}. 

Drowsiness can decrease cognitive performance including vigilance, information processing and decision making \cite{RN26}. Reaction time is also decreased under the influence of drowsiness, making it unsafe to drive. Visible symptoms can include frequent yawning, eye closure, head tilt, itchy or red eyes, lack of attention, line crossing, a tendency to accelerate, and reduced stopping distances \cite{RN25}. 

Reducing the effects of drowsiness can be achieved by taking a nap, while avoiding drowsiness in general can be achieved through regular sleep patterns (more than 7 hours a night) and avoidance of regular daytime napping \cite{RN27}. Research has shown that caffeine consumption is not beneficial in the long term and can induce sleep in the short term (1-2 hours) \cite{RN25}. However, slow-release caffeine can improve performance for over 5 hours \cite{RN28}. Moreover, regular caffeine intake can provide minimal effect on a user, hence, this should not be used as a drowsiness countermeasure. 

As drowsiness is difficult to define, subjective measures have been used as a label in most cases for data gathered in studies of fatigue, where subjects have given a number associated with their level of drowsiness. Alternatively a professional may label data based on visual drowsiness indicators. Numerical drowsiness scores are used for the majority of studies, with most studies using the Karolinska Sleepiness Scale (KSS) \cite{RN7}. Professional labelling has also been performed on video recordings of subjects, where an Observer Rating of Drowsiness (ORD) is often used. Less common methods are also discussed in this paper. 

The main methods explored to monitor drowsiness include physiological, behavioural, and vehicle-based technologies \cite{RN8}. Physiological methods that have been targeted are mainly based on measurements of brain waves, cardiovascular measurements, muscle fatigue, eye movement, and eye closure. These methods have been shown to be most accurate, however, these are also the most intrusive for measuring drowsiness  \cite{RN9, RN10, RN11}. Behavioural measures are monitored using a camera, where facial features, head inclination, yawning, and eye features are used to measure drowsiness \cite{RN10}. Behavioural methods are said to be the second most accurate and less intrusive than other methods \cite{RN12}, however they may be subject to privacy concerns \cite{RN140}. Finally, vehicle-based measures have focused mainly on lane departure and steering wheel measurements \cite{RN13, RN14}. 

Studies have suggested that multi-modal (physiological) and hybrid approaches produce greater accuracy compared to a mono-signal or single drowsiness detection method \cite{RN15, RN16, RN17, RN18, RN19, RN20, RN21}. Despite this, fewer studies have focused on hybrid models. Furthermore, commercial products mainly focus on behavioural and vehicle-based measures, with the latter being most commonly used in vehicle, as they are most easily implemented by automotive manufacturers. Physiological methods are not widely used commercially, perhaps due to their intrusiveness.

\section{Evaluation Metrics}
Throughout this paper we discuss various metrics used to help categorise and summarise the papers surveyed. To clarify our evaluation criteria, we define the following aspects to determine the benefits and downfalls of a system, which were areas of importance and concern in various papers:
\begin{itemize}
    \item Accuracy: the reported percentages of correct drowsy and non-drowsy classification
    \item Intrusiveness: both physical and psychological intrusiveness, which will be defined subsequently
    \item Privacy: the ability to identify people or sensitive information using collected data
    \item Data size: how much data will need to be stored 
    \item Cost: how much the acquisition system will cost to implement in a vehicle
    \item Susceptibility to noise: how much noise will be present in the raw data
    \item Vulnerability to subject diversity: how much the model will be affected by differences in ethnicity, age and gender.
    \item Adaptability: how well the system can cope with data loss
\end{itemize}

Our analysis of the literature showed that definitions of intrusiveness within papers could vary and that the levels of intrusiveness were not clear. Hence, we first introduce a new intrusiveness scale, in order to evaluate systems consistently. Then, based on the frequency of concern in papers of the identified areas, we rank the existing non-hybrid methods according to the dimensions above.

\subsection{Defining Intrusiveness}
In many of the papers surveyed the word "intrusive" was not directly defined, although in some instances, methods were described using phrasing such as "less intrusive". In order to clarify types and levels of intrusiveness, we propose a new scaling system as per the following:
\begin{itemize}
    \item Low: No intrusion and not noticeable by the user. For example, a sensor could be completely immersed within a car seat.
    \item Medium:
    \begin{itemize}
        \item Type I: Psychological intrusiveness where a subject is personally monitored by video or images and can be easily identified.
        \item Type II: Physical intrusiveness where minimal contact required, for example sensors that are noticeable and in contact but minimalist and lightweight.
    \end{itemize}
    \item High: Physically intruding on a person, sensors can be felt when movement occurs
\end{itemize}
Henceforth, our paper uses this proposed intrusiveness scale when referring to levels of intrusiveness.

\subsection{Ranking of current methods}
Table \ref{tab:ranking} provides a ranking of the various methods in the literature and how they perform for each category of interest discussed above, based on the frequency with which a given dimension was mentioned in the literature surveyed. In Table \ref{tab:ranking}, a "1" is used to denote the best performing method, while "3" denotes the worst performing method for a given category.

\section{Commercially Available Systems}
A number of commercial products have been implemented to counter drowsy driving, many of which use vehicle-based methods and are installed by car manufacturers. Other commercial devices include behavioural and physiological components, where the detection systems are optional external purchases (targeted at truck, mining and logistic companies). 

\subsubsection{Physiological Commercial Products}
Existing physiological products (4) can be relatively expensive, with 3 built around Galvanic Skin Response (GSR). One product includes heart rate and the last physiological product (SmartCap) is targeted at mining settings and uses EEG monitoring. These products are listed in Table \ref{tab:products}.

\begin{table}[!t]
\renewcommand{\arraystretch}{1.3}
\caption{Physiological Commercial Products}
\centering
\begin{tabular}{|c|c|c|}
\hline
\textbf{Product} & \textbf{Company} & \textbf{Product Notes} \\
\hline
Anti Sleep alarm Vigiton \cite{RN32} & Neurocom & GSR wristband and ring \\
\hline
StopSleep \cite{RN33} & StopSleep & EDA (GSR)\\
\hline
STEER \cite{RN34} & STEER & HR + EDA (GSR) \\
\hline
SmartCap \cite{RN35} & SmartCap & EEG \\
\hline
\end{tabular}
\label{tab:products}
\end{table}

\subsubsection{Behavioural Commercial Products}
Commercial products relying on behavioural sensing are the most commonly available product that can be purchased for detecting drowsiness \cite{RN36}. These make use of cameras focused on the driver and are generally based on eye tracking. The products identified are shown in Table \ref{tab:behavioural}.

\begin{table*}[!t]
\renewcommand{\arraystretch}{1.3}
\caption{Behavioural Commercial Products}
\centering
\begin{tabular}{|c|c|c|}
\hline
\textbf{Product} & \textbf{Company} & \textbf{Product Notes} \\
\hline
Delphi DSM \cite{RN37} & Delphi & Uses a near IR camera \\
\hline
Eagle light \cite{RN38} & Optalert & Uses glasses to detect blinking\\
\hline
AntiSleep \cite{RN39} & SmartEye & Eye and head tracking \\
\hline
EyeAlert \cite{RN40} & LumeWay & Eye closure rate and duration \\
\hline
Eyetracker System \cite{RN41} & Fraunhofer & Dual camera for eye tracking \\
\hline
Siemens IR-LED \cite{RN42} & Siemens & Detects microsleep \\
\hline
Operator alertness system light vehicle \cite{RN43} & HxGN (MineProject) & Eye and face tracking \\
\hline
Guardian \cite{RN44} & Seeing Machines & General camera tracking \\
\hline
Saab Driver Attention Warning System \cite{RN45} & SAAB & Uses 2 IR cameras to detect eyes, gaze and head pose\\
\hline
DADS \cite{RN46} & InterCore & Bluetooth to smartphone monitoring \\
\hline
DriveAlert+ \cite{RN47} & DriveRisk & Eyes, facial features and movement \\
\hline
Driver fatigue monitoring system \cite{RN48} & STONKAM® & General camera tracking \\
\hline
Driver fatigue monitor DFM2 \cite{RN49} & Transport Support & PERCLOS \\
\hline
\end{tabular}
\label{tab:behavioural}
\end{table*}

\subsubsection{Vehicle-based Commercial Products}
Table \ref{tab:vehicle} includes current companies, their vehicle-based drowsiness detection products and the methods they use to detect drowsiness.

\begin{table*}[!t]
\renewcommand{\arraystretch}{1.3}
\caption{Vehicle-Based Commercial Products}
\centering
\begin{tabular}{|c|c|c|}
\hline
\textbf{Product} & \textbf{Company} & \textbf{Product Notes} \\
\hline
FAS 100 \cite{RN50} & Albrecht & Lane assist, only for $>75$km/hr \\
\hline
Driver alert \cite{RN51} & Ford & Lane crossing \\
\hline
Attention Assist \cite{RN51} & Mercedes, Nissan, BMW & Steering wheel movement \\
\hline
Driver alert control system \cite{RN52} & Volvo & For less consistent driving \\
\hline
SafeTRAC \cite{RN53} & Cognex & Lane positioning \\
\hline
Safety System+  \cite{RN54} & Lexus & Lane departure warning \\
\hline
EyeSight driver assist \cite{RN55} & Subaru & Lane keeping and object detection \\
\hline
ASTiD \cite{RN56} & fmi & Time of day + SWM \\
\hline
Iteris Safety Direct \cite{RN57} & Iteris & Lane departure warning \\
\hline
Rest Assist \cite{RN58} & Volkswagen, Skoda & Steering Wheel Movement (SWM) \\
\hline

\end{tabular}
\label{tab:vehicle}
\end{table*}

\subsubsection{Hybrid Products}
We identified two implementations of hybrid methods in cars, including Toyota drowsiness tracking \cite{RN30}, and the Bosch mobility solutions Interior Monitoring System \cite{RN31}. Both hybrid systems do not include physiological components, instead combining vehicle and behavioural approaches. 

Current commercial products are largely reliant on vehicle-based methods, which have shown to be the least accurate and reliable for detecting drowsiness, although this is the easiest data to gather and process. These systems are not always defined as drowsiness devices, but also targeted at distraction detection. They are often labelled in vehicles as lane departure and lane assist products. More reliable and specific devices are required for better accuracy, including the implementation of more complex systems. Moreover, commercial products are focused on detection rather than prediction which is a major disadvantage of current devices. Hence, further research and development is required.

\section{Methods under active research and development} 
For the most part the development of drowsiness detection and prediction technologies has followed a standard paradigm. First, labelling of drowsiness has been undertaken using subjective measures, which is then used to develop models in one of the four primary categories: physiological, behavioural, vehicle-based, and hybrid.

\subsection{Drowsiness Labelling}
Subjective measures are often used to describe a subject’s state of drowsiness. The most common subjective method used in the literature is the Karolinska Sleepiness Scale (KSS), followed by the Stanford Sleep Scale (SSS) \cite{RN7}. Other sleep scales seen in the literature include the HFC drowsiness scale, Epworth Sleepiness Scale (ESS), Johns Drowsiness Scale (JDS), Observer rating of drowsiness (ORD), and Subjective Drowsiness Rating (SDR). The HFC drowsiness scale, ORD rating and SDR are externally labelled, the JDS rating is a physiological indication and the remainder of labelling methods are self estimations. 

\subsubsection{Karolinska Sleepiness Scale (KSS)}
\begin{figure*}[!t]
\centering
\begin{overpic}[clip, trim=0.7cm 18cm 2cm 2.4cm, width=6.7in]{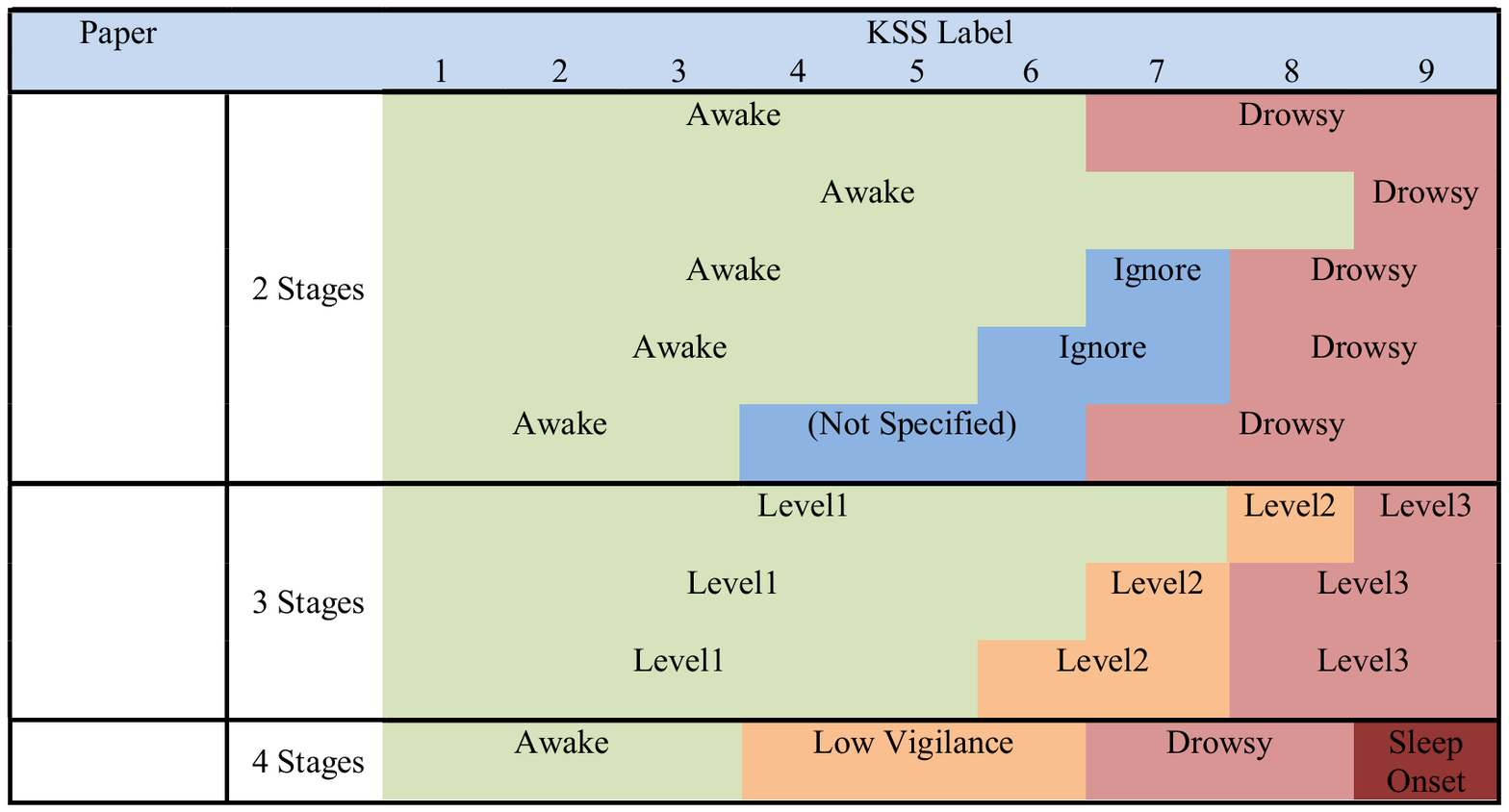}
\put(10,104){{\parbox{0.15\linewidth}{%
     \cite{RN59, RN60} \\
     \\
     \cite{RN61} \\
     \\
     \cite{RN62} \\
     \\
     \cite{RN63} \\
     \\
     \cite{RN64} \\
     \\
     \cite{RN61} \\
     \\
     \cite{RN65} \\
     \\
     \cite{RN59, RN66, RN67} \\
     \\
     \cite{RN68}
     }}}
\end{overpic}
\caption{KSS drowsiness labelling used in current models.}
\label{fig:KSS}
\end{figure*}

The KSS was the most used subjective scale in the literature, following a 9-point system (Table \ref{tab:KSS}). Some studies used the KSS to categorise fatigue into two stages (awake and drowsy), others three stages (awake, slight and drowsy), and sometimes 4 stages. Of those studies simplifying to two stages, many left out KSS points to enable a clear difference between awake and drowsy states. The use of KSS in evaluating fatigue is shown in Fig. \ref{fig:KSS}.

\begin{table}[!t]
\renewcommand{\arraystretch}{1.3}

\caption{KSS Scoring System}
\label{tab:KSS}
\centering

\begin{tabular}{|c||c|}
\hline
Extremely alert & 1\\
\hline
Very alert & 2\\
\hline
Alert & 3\\
\hline
Rather alert & 4\\
\hline
Neither alert nor sleepy & 5\\
\hline
Some signs of sleepiness & 6\\
\hline
Sleepy, but no difficulty remaining awake & 7\\
\hline
Sleepy, some effort to keep alert & 8\\
\hline
Extremely sleepy, fighting sleep & 9\\
\hline
\end{tabular}
\end{table}

\subsubsection{Stanford Sleep Scale (SSS)}
\begin{figure*}[!t]
\centering
\includegraphics[clip, trim=2.34cm 25cm 2cm 2.4cm, width=6.55in]{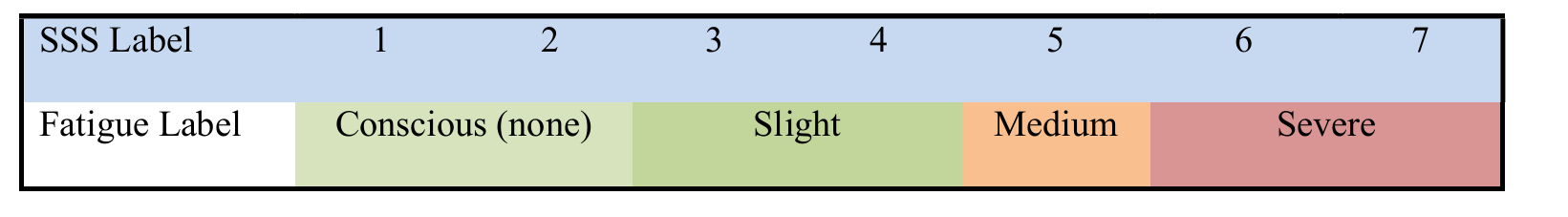}
\caption{SSS drowsiness labelling used in current models.}
\label{fig:SSS}
\end{figure*}

The Stanford Sleep Scale (SSS) (Table \ref{tab:SSS}) has been used similarly to the KSS scale, with one study in particular grouping the ratings to reduce the number of output labels \cite{RN69}. The SSS is a 7-point scale, where one proposed combination of stages for fatigue labelling is shown in Fig. \ref{fig:SSS} \cite{RN69}. 

\begin{table}[!t]
\renewcommand{\arraystretch}{1.3}
\caption{SSS Scoring System}
\label{tab:SSS}
\centering
\begin{tabular}{|c||c|}
\hline
Wide awake / alert & 1\\
\hline
Functioning at high levels but not fully alert & 2\\
\hline
Awake, relaxed, responsive but not fully alert & 3\\
\hline
Somewhat foggy & 4\\
\hline
Foggy, slowed down & 5\\
\hline
Sleepy, woozy, fighting sleep & 6\\
\hline
No longer fighting sleep, sleep onset soon & 7\\
\hline
Asleep & -\\
\hline
\end{tabular}
\end{table}

\subsubsection{Epworth Sleepiness Scale (ESS)}
The ESS is used to measure daytime sleepiness with a questionnaire on the likelihood of dozing during eight common situations \cite{RN70}. These situations include dozing when watching TV, as a passenger in a car, talking to someone, sitting in traffic etc. The subject is asked to rate their chance of dozing based on the following options,
\begin{itemize}
    \item Never doze = 0
    \item Slight chance of dozing = 1
    \item Moderate chance of dozing = 2
    \item High chance of dozing = 3
\end{itemize}
where a score between 0 and 9 are labelled to be awake \cite{RN25}. This method is useful for an overall view of tendency to sleep in subjects, but cannot be used for detection \cite{RN63}.

\subsubsection{The Groningen Sleep Quality Scale (GSQS)}
Similarly to the ESS, the GSQS comprises a pre-study questionnaire that is used to measure the quality of sleep the night before. The GSQS was used in one study to determine the continuation of the experiment with a particular subject, where exceeding the sleep quality threshold of 3 meant the subject could not participate, as that indicated they had intermediate sleep disturbances \cite{RN71}. The GSQS asks 15 true or false questions, with a maximum score of 14 available. 

\subsubsection{The Chalder Fatigue Scale (CFS/CFQ)}
The CFQ (adapted from the CFS) is an 11-question survey aiming to measure fatigue levels, with reference to a “usual” state \cite{RN72}. The former CFS scale comprises 14 questions, where a yes or no answer is required for each \cite{RN73}. A study showed that the CFS showed fatigue levels similar to that of the more commonly used SSS \cite{RN71}.

\subsubsection{HFC Drowsiness Scale}
The HFC Drowsiness Scale (Table \ref{tab:HFC}) is used to annotate video-based recordings, where trained annotators label video recordings every minute \cite{RN67}.

\begin{table}[!t]
\renewcommand{\arraystretch}{1.3}
\caption{HFC Drowsiness Scale}
\centering
\begin{tabular}{|c||c|}
\hline
Wide awake, vivid attention & 1\\
\hline
Highly concentrated, focused attention & 2\\
\hline
Attentive but calm & 3\\
\hline
Neither activated nor drowsy & 4\\
\hline
Somewhat dozy but ready to respond & 5\\
\hline
Signs of drowsiness, but effortlessly awake & 6\\
\hline
Clearly drowsy, but focused on driving task & 7\\
\hline
Fight against drowsiness, but largely perceptive & 8\\
\hline
Absent minded, long periods without activity & 9\\
\hline
\end{tabular}
\label{tab:HFC}
\end{table}

\subsubsection{Observer rating of drowsiness (ORD)}
\begin{figure*}[!t]
\centering
\begin{overpic}[clip, trim=2.4cm 24cm 2cm 2.4cm, width=7in]{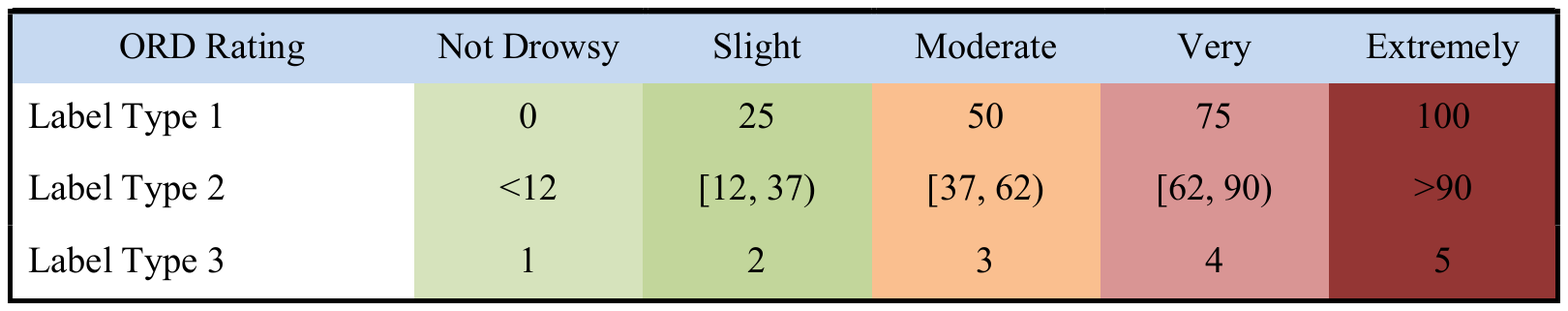}
\put(80,30){{\parbox{0.1\linewidth}{%
     \cite{RN74} \\
     \\
     \cite{RN75} \\
     \\
     \cite{RN76, RN77} \\
     }}}
\end{overpic}
\caption{ORD in current models.}
\label{fig:ORD}
\end{figure*}

Two different levels of ORD have been used, both for the analysis of video recordings, based on the face and body language of a driver \cite{RN74}. The first ORD is 100-point rating scale where two studies have grouped the ratings into drowsiness labels. The second ORD is a 5-point scale, with the same classifications as the first ORD but labelled under the numbers 1 to 5. A summary can be found in Fig. \ref{fig:ORD}, where one study grouped moderately, very and extremely drowsy into a single “drowsy” label for classification \cite{RN75}.

\subsubsection{Subjective Drowsiness Rating (SDR)}
The SDR was developed for a study that used a 5-point system, very similar to the 5-point ORD \cite{RN78}. The fatigue stages used in the study were based on behavioural signs observed in a video recording, where a non-drowsy to extremely drowsy rating was indicated using the labels 0 to 4. 

\subsubsection{Johns Drowsiness Scale (JDS)}
The JDS is an objective labelling system for drowsiness. It is based on reflectance oculography measures including relative velocity of blinks, relative duration of blinks as well as other eye closure measures \cite{RN79}. JDS is measured using the Optalert device and is a 10-point scale where below 4.5 is alert, 4.5 to 5 is considered drowsy and critical drowsiness is anything above a rating of 5 \cite{RN80}.

\subsection{Physiological Methods}
Approaches to driver drowsiness detection and prediction have been undertaken using both uni- and multi-modal methods, with multi-modal approaches proving more reliable \cite{RN15}. Acquiring these signals can require a high level of intrusiveness, negatively impacting their potential use in detecting driver drowsiness \cite{RN9}. Alternatively, some of these signals can be collected using medium to low levels of intrusiveness, with sensors placed in the steering wheel of the vehicle \cite{RN81, RN82}, drivers seat \cite{RN11, RN83} or by using wrist worn sensors \cite{RN9}. Physiological signals have shown to be more accurate when compared to vehicle-based and behavioural methods as they are less prone to interference from lighting, environmental and road conditions \cite{RN10, RN11}. Furthermore, physiological signals can detect drowsiness at earlier stages when compared to behavioural and vehicle methods \cite{RN10}.

\begin{figure*}[!t]
\centering
\includegraphics[clip, trim=2.5cm 1cm 2.5cm 0cm, width=6.4in]{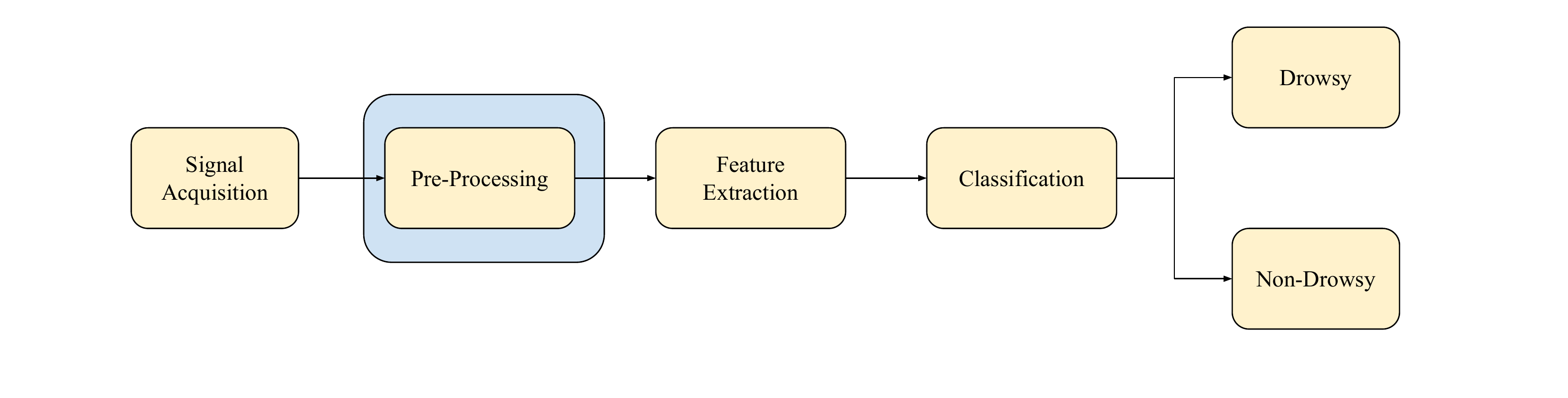}
\caption{Process for physiological signals to determine driver drowsiness.}
\label{fig:Phys}
\end{figure*}

The main physiological measurement methods include  Electroencephalography (EEG) \cite{RN156, RN157, RN159}, Electrocardiography (ECG) \cite{RN155, RN161, RN154}, Electromyography (EMG) \cite{RN77, RN92} and Electrooculography (EOG) \cite{RN127, RN19, RN160}; moreover, some studies have also explored Galvanic Skin Response (GSR) \cite{RN92, RN18}, Ballistocardiography (BCG) \cite{RN150}, Seismocardiography (SCG) \cite{RN150} and Skin Temperature (ST) \cite{RN21}. Photoplethsmogram (PPG) has also been applied with similar drowsiness correlations to ECG \cite{RN17, RN158, RN162}. These signals are typically processed and analysed for use in predicting drowsiness using the process shown in Fig. \ref{fig:Phys}.

Certain signals are able to be obtained with lower levels of intrusiveness than others. Data capture mechanisms and corresponding levels of intrusiveness are described in more detail below. Note that the relationship of drowsiness to certain features is not elaborated upon here, as the following previously mentioned review papers provide a thorough overview of certain features and their relationship to drowsiness \cite{RN10, RN279, RN36, RN11, RN7, RN29}.
 
 \subsubsection{EEG} 
 For full EEG data collection, a cap with a large number of electrodes is often used, where the most relevant data can be collected from the occipital lobe at the back of the head \cite{RN8, RN84}. Traditional EEG has a high level of intrusion, but alternative methods specific to driver drowsiness detection including the use of a headband \cite{RN85, RN86} and in-ear EEG have been used \cite{RN18}. Nonetheless, EEG signal acquisition faces a trade-off between intrusiveness and accuracy, where systems offering lower levels of intrusion, such as headband EEG recording, provide a lower detection accuracy when compared to conventional methods \cite{RN86}.
 
\subsubsection{Heart Rate monitoring methods}
ECG also has a high level of intrusiveness, but can be collected less intrusively by using electrodes in a steering wheel rather than the chest \cite{RN82}. However, this requires direct contact with the steering wheel using both hands. PPG can be used as an alternative to ECG with medium level intrusiveness and is an easy to collect signal \cite{RN17, RN18}. Furthermore, BCG and SCG do not require skin contact and offer a low level of intrusiveness \cite{RN150}. These methods can use sensors positioned in inconspicuous locations including the head rest of a driver's seat \cite{RN83, RN95, RN150}, the car's steering wheel \cite{RN5, RN96, RN150}, the seat belt \cite{RN150} or on the finger \cite{RN17}. Furthermore, video monitoring can also be used to detect heart rate and heart rate variability in both day and night conditions, eliminating the need for sensors \cite{RN151, RN152}. However, the alternative methods are more susceptible to noise and data loss. The steering wheel sensor is not viable when the subject is using gloves and these methods are more susceptible to movement noise than alternative capture methods. 

\subsubsection{EOG}
EOG recordings require electrodes to be placed close to the eyes for a stronger signal \cite{RN63, RN87}. This associates EOG recordings with a high level of intrusion; however, studies have suggested this can be replaced with a camera for many of the blink features typically identified using EOG \cite{RN19, RN88}. This shifts the intrusiveness of acquisition to psychological intrusiveness, corresponding to a medium level of type I intrusiveness. 

\subsubsection{EMG}
EMG recordings are often captured using surface EMG (sEMG) as this is a non-invasive method of collecting EMG signals. Electrodes have been placed on the legs (anterior tibialis) \cite{RN89}, chin \cite{RN89}, upper arm (on the deltoids) \cite{RN90, RN91}, back (on the trapezius) \cite{RN90, RN91}, neck (splenius capitis) \cite{RN91} as well as lower arm \cite{RN92}. EMG acquisition is associated with high-level of intrusion. 

\subsubsection{GSR}
The GSR signal acquisition can be categorised as a medium or a low level intrusiveness as it can be placed in a steering wheel, but contact may still required if used in wearable devices \cite{RN18, RN92, RN93}. 

\subsubsection{Temperature}
Temperature that has been taken externally at the forehead was shown to have the greatest relationship to drowsiness, when compared to body temperature captured at other locations \cite{RN94}. Other points for detecting skin temperature have been proposed through the steering wheel alongside various other methods, both with medium and low levels of intrusion \cite{RN82}.

\subsubsection{Breathing}
Respiration sensors offer a medium level of intrusion, for example those using a band around the chest \cite{RN17}. Respiration can also be measured from videos, making the process non-contact and of a low level of intrusion \cite{RN151, RN152}. However, this method can be challenging due to noise and may be difficult to apply on road. A hybrid method can be used in this case, where the collection methods can be cross-validated between a seat belt, camera, heart-rate derived breathing and seat pressure to determine the true breathing rate of a subject and help counter the effects of noise.

\subsubsection{Thermal Imaging}
Thermal imaging has been proposed to remove all physical intrusiveness, and is categorised as a method with a low level of intrusiveness, where it is a viable method in detecting drowsiness \cite{RN76}.

\subsubsection{Grip Strength}
Grip strength is another feature that can be integrated into the steering wheel, which has been correlated with EMG in predicting muscle fatigue \cite{RN97}. Grip strength data collection can also offer low level intrusiveness as this can be integrated in the steering wheel.

Many of the features described above can be combined or extracted using other acquisition systems to reduce the number of sensors required to collect relevant drowsiness features. For example, respiration (thermal imaging and respiration sensors) can be derived from ECG or video, reducing the need for a breathing belt. Low and medium intrusive alternatives could also be used instead of traditional methods. PPG and BCG have a  medium intrusive level, whereas ECG has a high level of intrusiveness, but can extract some of the same features. These methods can be useful in reducing the cost and intrusiveness of signal acquisition.

\subsection{Behavioural Methods}
Driver behaviour monitoring is becoming more popular as it is not a physically intrusive approach \cite{RN98}; hence proving more desirable than physiological monitoring. However, behavioural methods are psychologically intrusive, due to their association with surveillance, falling under the medium type II category for intrusiveness. Behavioural methods have been shown to be more accurate than vehicle monitoring \cite{RN12} and uses cameras to detect drowsiness indicators such as yawning, eye features, facial features and head orientation \cite{RN10}. Eye feature detection can be used as a non-invasive alternative of EOG for measuring blink duration, blink frequency and percentage of eyelid closure (PERCLOS), where the general measure is where the eye is 80\% closed. Of the behavioural indicators, two studies reported that head movement/pose features provide the highest correlation with drowsiness \cite{RN98, RN99}; however, eye movement-based measures are the most widely used measures \cite{RN7}.

\begin{figure*}[!t]
\centering
\includegraphics[clip, trim=1cm 0cm 1cm 0cm, width=7in]{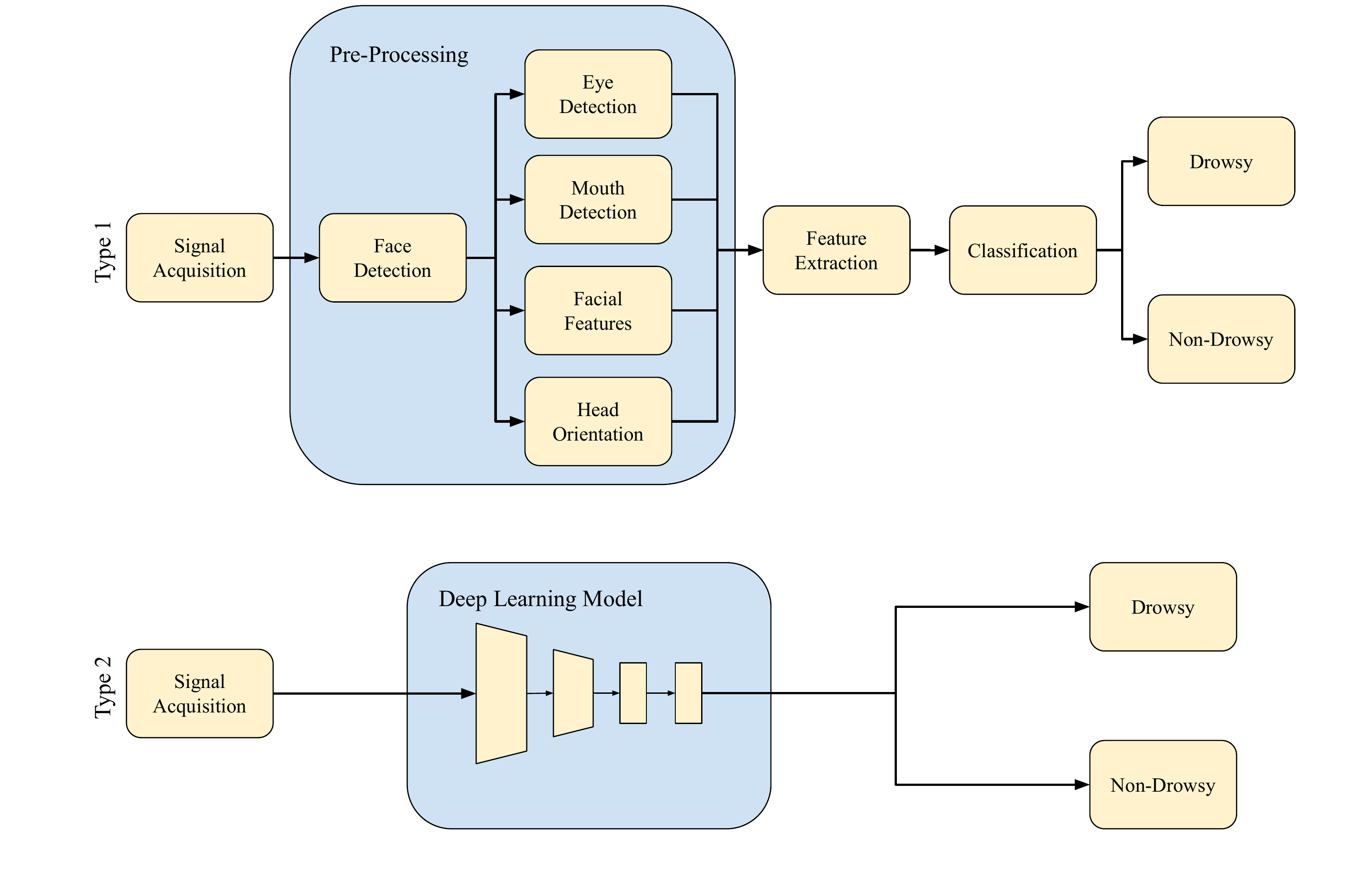}
\caption{Process for behavioural driver drowsiness detection.}
\label{fig:Behav}
\end{figure*}

Behavioural detection processes tend to follow physiological drowsiness detection methods, involving collecting, pre-processing, extracting features, and then classifying the data into drowsy and non-drowsy states. The pre-processing phase of behavioural methods can involve face detection, followed by detecting the mouth, eyes, facial features and head pose. However, often the processing of behavioural data involves end-to-end processes uses a deep learning model. The two types of processes are shown in Fig. \ref{fig:Behav}, where Type 2 involves end-to-end processing.

\subsection{Vehicle-Based Methods}
Vehicle based methods are the least common and least accurate method for detecting driver drowsiness, and most suited to roads with clean line makings and good weather conditions \cite{RN12}. However, these methods are the simplest to implement, and most convenient when it comes to data collection. They use lane position, steering wheel, acceleration pedal, and yaw features to evaluate driver drowsiness \cite{RN13, RN14}. Steering wheel parameters were found to be the most commonly investigated in recent literature despite one study suggesting lane keeping parameters were the most widely used approach \cite{RN7}. One study showed that steering wheel parameters were more accurate than lane parameters \cite{RN100}; however, another study suggested that lane variability was better \cite{RN14}. 

\begin{figure*}[!t]
\centering
\includegraphics[clip, trim=2.5cm 1cm 2.5cm 0cm, width=6.4in]{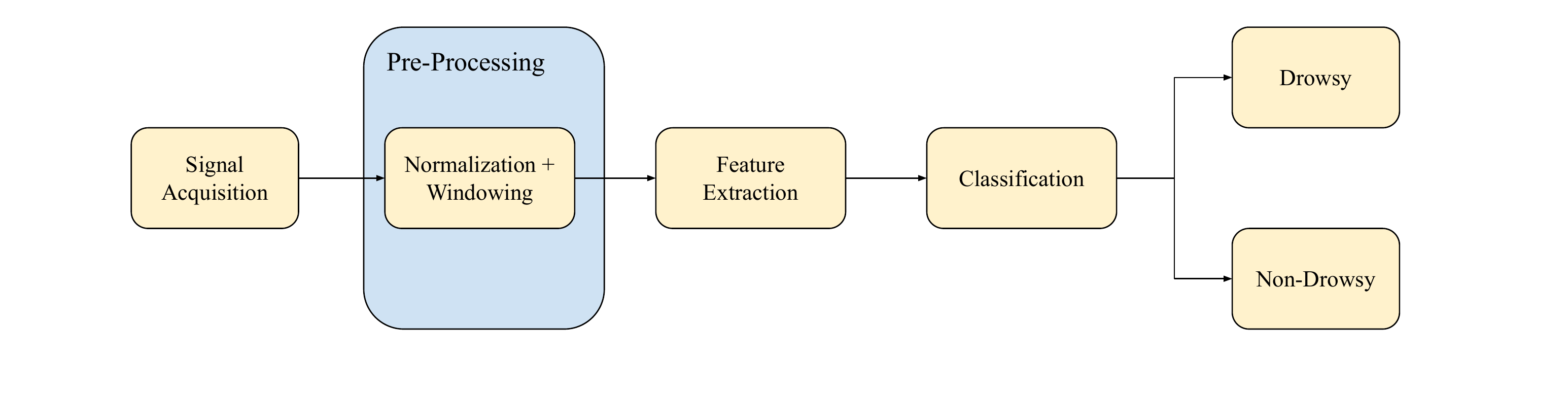}
\caption{Flow chart of vehicle-based detection.}
\label{fig:Veh}
\end{figure*}

The recording and modelling of vehicle-based methods is simpler than behavioural and physiological methods. The vast majority of studies described little pre-processing other than normalization or standardization and windowing of the data. Hence, vehicle-based drowsiness detection follows the pattern described in Fig. \ref{fig:Veh}.

\subsection{Hybrid Methods}
Hybrid methods are relatively new in the literature as the individual methods had to be established before the combined approaches \cite{RN101}. Thus, most  studies of hybrid technologies have reported better results \cite{RN16, RN17, RN18, RN19, RN20, RN21}, indicating that hybrid approaches are the best method for detecting drowsiness in terms of lowering intrusiveness levels, accuracy, and ability to function with data loss. This is due to a combination of different techniques being able to help overcome the negative components of specific techniques \cite{RN11}. 

Hybrid models use the same features as the physiological, behavioural, and vehicle-based methods, with individual models often being developed for each method and combined to develop the hybrid model. Hence, the data is often classified using the individual method, then again with other parameters considered. Additional methods are also sometimes present in hybrid models, such as movement \cite{RN102} or seat pressure detection \cite{RN103, RN104, RN105}, driving duration \cite{RN64, RN106} and voice input \cite{RN107}. Furthermore, other features appeared in hybrid methods that weren't common across individual method studies including gaze direction and pupil diameter in behavioural methods, vehicle speed, acceleration, and pedal input in vehicle methods. It was found that hybrid studies are able to minimise limitations seen in the other methods, however, some still exist and will be discussed in-depth in the following sections as these are the current limitations in progressing driver drowsiness technology.

\section{Existing Challenges}

\subsection{Detection versus Prediction}
Behavioural and vehicle-based methods focus on the detection of drowsiness rather than prediction, despite the label “prediction” being used \cite{RN106}. For behavioural methods, often prolonged eye closure is an indicator of drowsiness, where a microsleep may have already occurred. Furthermore, the commonly used measure, PERCLOS, has been found to detect drowsiness too late \cite{RN87}. In vehicle-based studies, often lane departure is an indication of drowsiness; however, once a car has departed the lane, it has the potential to crash into oncoming traffic, indicating that this method is also too late to assist in driver safety. Advances in physiological monitoring has allowed for earlier prediction of drowsiness, where some features including changes in the VLF component of ECG can predict drowsiness up to 5 minutes before sleep \cite{RN89, RN108}. We recommend a time to event-based approach for vehicle-based and behavioural methods, where models are built to predict how long until the prolonged eye closure occurs, and focus on early prediction based on other features, rather than the detection of eye closure itself and the same for lane departure monitoring.  

\subsection{Feature Trend Discrepancies}
Hundreds of features have been explored across physiological, behavioural, and vehicle-based studies. Some studies have differing reports regarding particular features and their correlation with drowsiness. We caution the use of these, as these features could vary depending on acquisition, environment, processing or if incorrectly reported. Certain features may also be inherently different between subjects in different trials. Some of these features, in relation to increased drowsiness, include:
\begin{itemize}
     \item Breathing rate, where one study noted that breathing rate was assumed to decrease \cite{RN109} with drowsiness and another used thermal imaging to show this \cite{RN76}; however, no change was observed in another study \cite{RN89}. In contrast, one study reported there was an increase in breathing rate, where a breathing belt was used \cite{RN17}.
     \item The power of the low frequency (LF) component (ECG) was said to increase with drowsiness in \cite{RN108}; however, other studies found that it decreases \cite{RN89, RN110}.
     \item Blink frequency was said to decrease slightly in both \cite{RN111, RN112}; however, other studies suggested that blink frequency increases \cite{RN68, RN113}. Caffier et al. \cite{RN111} expands upon other studies and how blink frequency was found to be conflicting for different studies, where the variations could be due to differing circumstances and/or environments for the subjects.
\end{itemize}

Furthermore, other features in physiological detection have shown to differ before drowsiness and again once the subject is asleep. These can include: 
\begin{itemize}
     \item EEG: The spectral power of the alpha band, which increases before falling asleep then flattens afterward \cite{RN8}. 
     \item ECG: The power of the Low Frequency/High frequency component reduces before sleep then increases afterwards \cite{RN108}.
\end{itemize}

\subsection{Physiological Intrusiveness}
Physiological signals are considered be more intrusive, in particular when EOG and EEG signals are involved. The application of hybrid methods can address this, where a camera can be used to detect blinking parameters rather than EOG electrodes. Despite this, some studies still use glasses to monitor eye features \cite{RN64} and EEG/EOG recordings \cite{RN59, RN102, RN103, RN105, RN114, RN115}. This includes the Muse headband, which is less intrusive than traditional EEG, but still more intrusive than other methods \cite{RN86}. Accuracy reports have still been high however, when all medium level intrusive methods have been used.

\subsection{Subject Diversity}
Subject diversity has been lacking between subjects in areas including ethnicity, age, gender and glasses vs non-glasses users. Ethnicity variations can affect the accuracy of behavioural-based models, with most studies using just one ethnicity; however, a further problem exists where ethnicity is often not reported in studies. Research has shown that classifiers often reduce accuracy in darker skin tones, in particular for women, as models are often not diversely trained \cite{RN116}. Of the available datasets, the NTHU dataset is the most diverse, and well balanced \cite{RN117}; but, has still been flagged as a potentially problematic when it comes to generalisation for end-to-end learning models, which require large amounts of training data \cite{RN164}. This is a particular challenge in machine learning, as highlighted by the seminal gender shades  paper \cite{RN116}, that is yet to be adequately addressed in Drowsiness literature.

\begin{figure}[!t]
\centering
\includegraphics[clip, trim=1.1cm 0.2cm 0.2cm 0.2cm, width=3.4in]{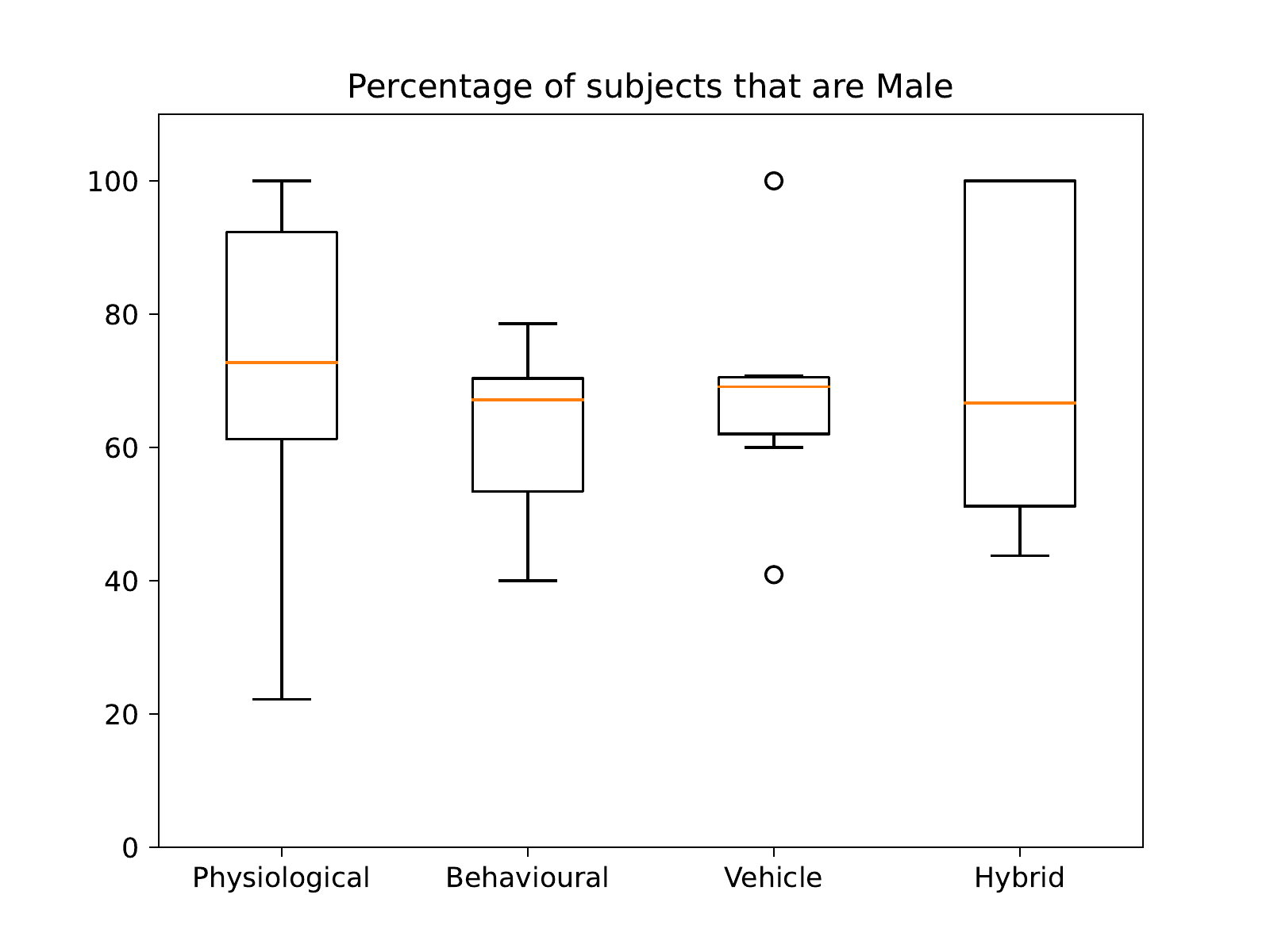}
\caption{Percentage of Male Subjects used in studies.}
\label{fig:Subs}
\end{figure}

Gender bias was largely present in the data with one study recognising this as biased \cite{RN103}. Fig. \ref{fig:Subs} demonstrates the percentages of male subjects in the studies, where 52 out of 126 studies reported the split of female and male subjects. Gender split is particularly important in behavioural and hybrid studies, as facial features can be different across gender.

\begin{figure}[!t]
\centering
\includegraphics[clip, trim=1.1cm 0.2cm 0.2cm 0.2cm, width=3.4in]{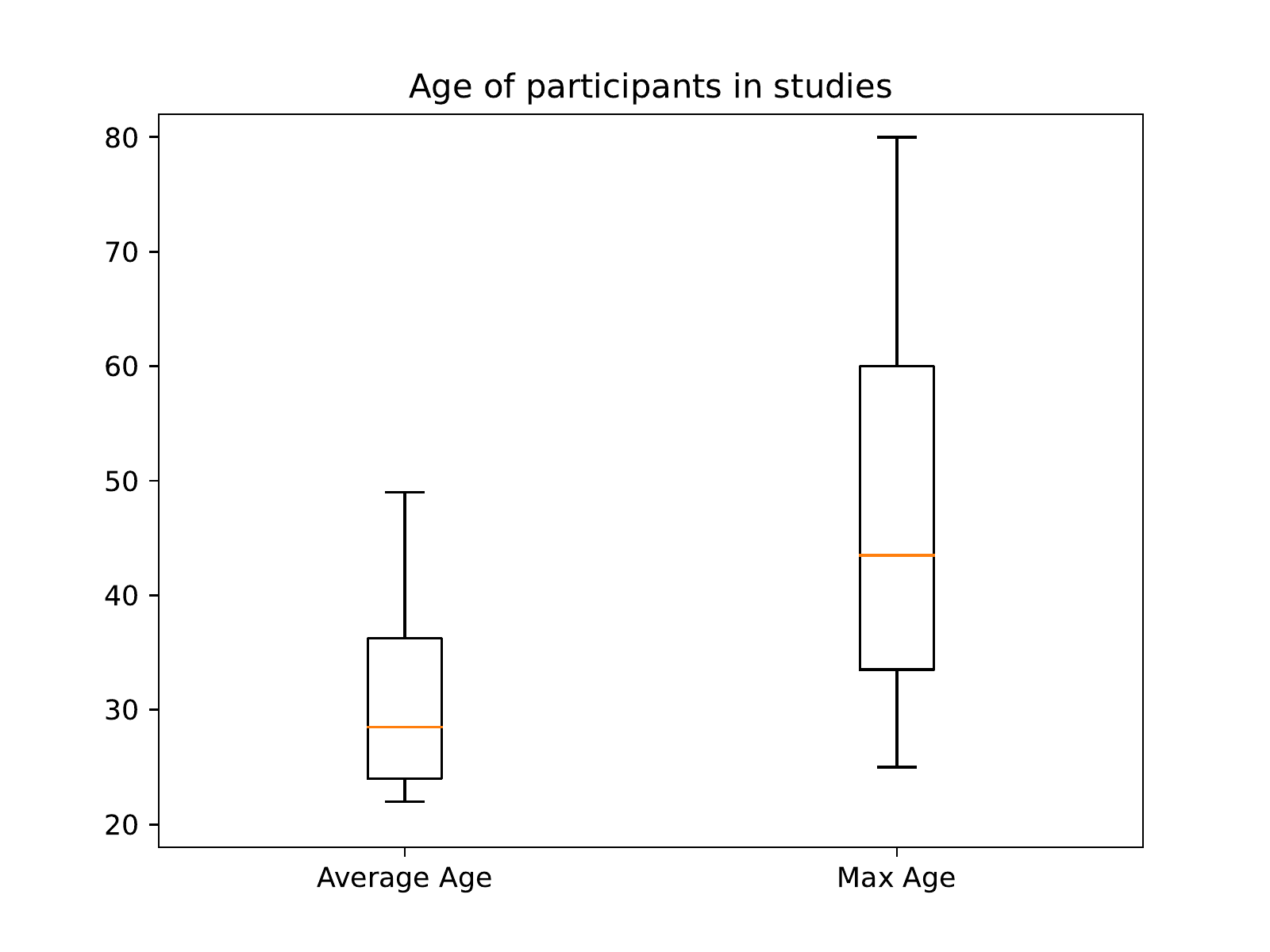}
\caption{Reported average and maximum ages of participants used in studies.}
\label{fig:age}
\end{figure}

Variance in ages of subjects is important when developing methods, as models can behave differently between younger and older participants. One hybrid study explored this, where age groups were split to compare those above and below 40 and a difference between age groups was found, where the model worked better for older subjects \cite{RN118}. In contrast, another behavioural study reported their model to be worse for older subjects \cite{RN119}, demonstrating that a more diverse range of age groups needs to be considered to allow and test for these discrepancies. However, as demonstrated in Fig. \ref{fig:age}, the median average age of participants is around 28-29, biasing models towards younger participants (average age reported in 24 studies). The maximum ages used in studies was more often reported (46 studies), where ages up to 80 were used; however, a concerning number of studies had the oldest participants of 40 or less (22 studies), as shown in Fig. \ref{fig:age}. 

Prescription glasses are worn full-time by 37\% of Australians, with 66\% of Australians reported wearing prescription glasses in general \cite{RN120}. Hence, glasses and sunglasses are regularly worn driving and studies have reported reduced accuracy levels in behavioural monitoring for glasses users \cite{RN119, RN121, RN122, RN123, RN124, RN125, RN126}. Despite this, behavioural studies and hybrid studies that include behavioural monitoring have specifically been reported to include glasses in only 47\% of studies. 

\subsection{Number of Subjects}
\begin{figure}[!t]
\centering
\includegraphics[clip, trim=1cm 0.2cm 0.2cm 0.2cm, width=3.4in]{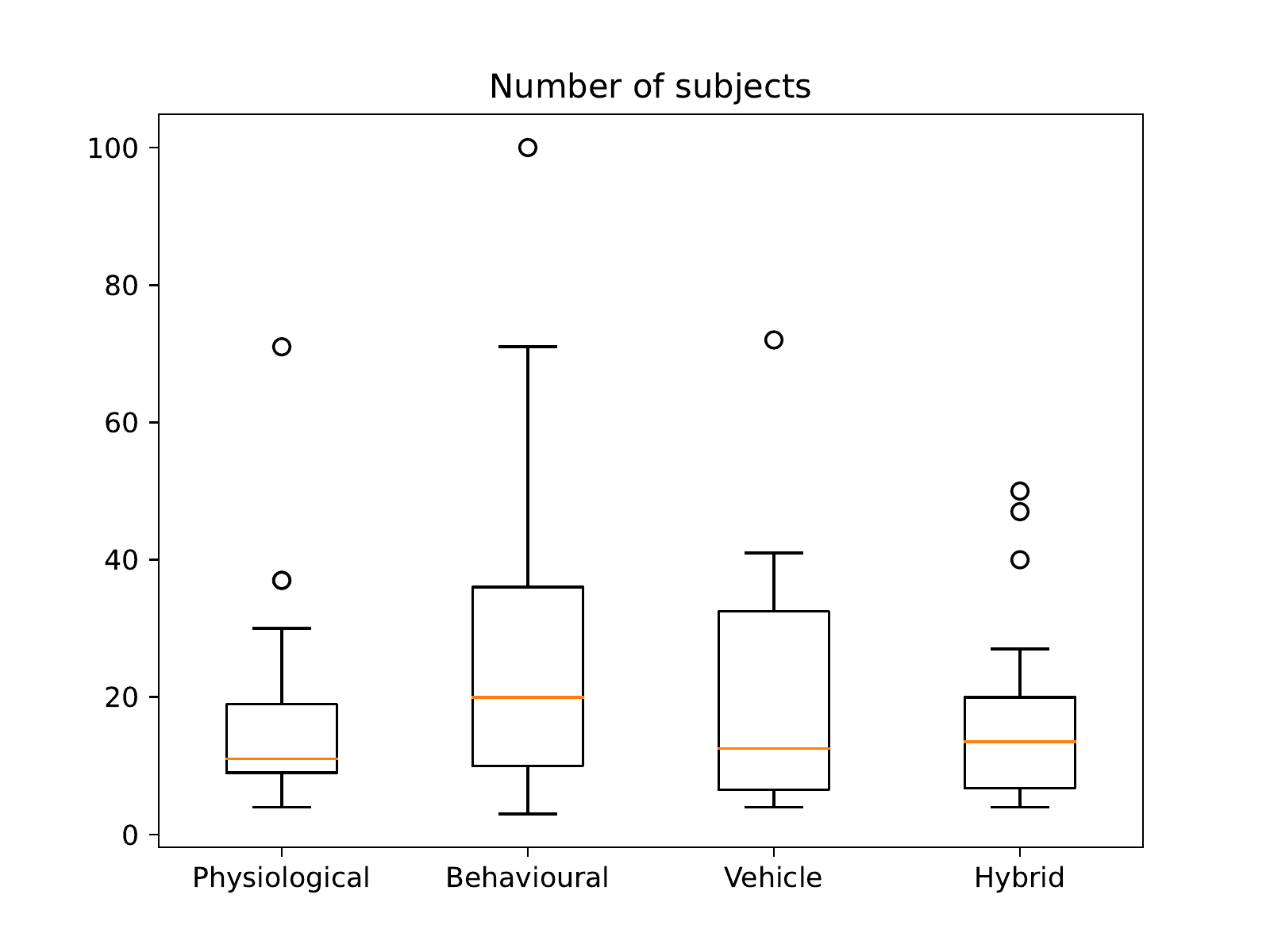}
\caption{Number of Participants in studies.}
\label{fig:Num}
\end{figure}

A limited number of subjects was repeatedly seen within the literature, where the number of subjects was often below 20 in studies, and a median of less than 15 seen in physiological, vehicle, and hybrid studies (Fig. \ref{fig:Num} ). A lack of subjects relates to the lack of diversity and models developed with such low numbers of participants cannot be appropriately generalised to the broader public. 

\subsection{Labelling of Drowsiness}
A variety of drowsiness labels have been used including the previously described subjective measures, EEG recordings, eye closure features, professional labelling, and lane departure occurrences. The different labels can affect what is considered drowsiness and hence produce different results between models. 

The most common measure for labelling drowsiness was the subjective measure using KSS \cite{RN7}; however, it has been noted that asking the driver can influence their drowsiness as the measure involves a verbal cue, where the driver is required to think \cite{RN66}. One study carried out experiments on collecting KSS ratings at times of 5, 10 and 15 minutes, where 10 minutes was found to be optimum \cite{RN66}.

Other methods such as professional labelling can vary depending on the professional's training; however, this measure can reduce the limitations of KSS measures. Lane departure, eye closure, and EEG measures have been used as both predictors and labels, where they do not have 100\% accuracy and could be an untrustworthy label. Lane departure and eye closure have also been shown to predict drowsiness too late, meaning these cannot be used as features to create a prediction model. EEG could be a suitable measure and provide consistency if a protocol was developed, as this is a more accurate measure and properly shows the subjects physiological state. However, this could also be influenced by equipment and artefacts, especially on the road. Hence, the labelling of drowsiness is not consistent across studies and remains a barrier to the implementation of driver drowsiness detection and prediction schemes.

\subsection{Accuracy reporting}
Accuracy measurements across studies are very difficult to compare. There are a variety of factors that need to be considered when comparing studies, hence it is not possible to compare studies without looking at the full set of parameters. Factors contributing to the accuracy of studies can include: 
\begin{itemize}
     \item Training, testing, and validation splits where the percentage split has been shown to affect models, in particular when a lower number of subjects are used. One study showed such split-related issues, where a higher accuracy was presented for a lower amount of testing subjects (80\% training, 20\% testing) \cite{RN127}. The most common reported and recommended split was 70\% training and 30\% testing.
     \item The level of drowsiness used – detecting slight drowsiness often provided a lower accuracy than severe drowsiness. 
     \item The number of drowsiness levels considered. Studies with more than 2 levels of drowsiness often produced lower reported accuracies. 
     \item It was observed that studies sometimes made models and collected baselines per subject which reported better accuracies than across subjects \cite{RN102}. Furthermore, studies that used cross-subject models should ideally have separate subjects in testing and training; however, they often used portions of each subject's data in both test and training sets, which does not validate the model for external subjects. 
     \item The number of subjects and diversity of subjects as previously discussed can also affect the reported accuracy of a study.
     \item Almost all studies use different datasets where the data they use or collect is not publicly available
     \item On-road and simulation studies will produce different results due to the different artefacts present in the car on-road. 
\end{itemize}

The numerical percentage for accuracy is often also expressed differently between studies, where some present the one value under “accuracy”, whereas others include specificity, sensitivity, precision, recall, root mean square error (RMSE) etc. Hence, it is difficult to compare accuracies across subjects, which is a particular concern when many researchers suggest that there are many other factors contributing to the outcome, including acquisition configuration, lighting and environment. 

\subsection{Timing}
The timing of feature extraction, batch processing, and duration of the experiment are all important factors when monitoring drowsiness. In some studies, the length of time used for extracting features was shown to exceed multiple minutes, which is not feasible for early detection. Longer window frames can run the risk that an adverse event has already occurred by the time of detection. Furthermore, some studies have collected chunks of data and labelled the entire segment as drowsy or non-drowsy; however, drowsiness can change at any given time. Therefore, the collected data may not be a true representation of the instantaneous drowsiness of a subject and can alter a models’ perception of drowsiness. Experiment duration should also be considered as some studies have reported links between duration of drive and drowsiness \cite{RN106}. 

\subsection{Data Collection Protocol}
Within the literature, it was found that experiment protocols need improvement and consistency \cite{RN7}. On top of the previously discussed issues, protocols need to include both day and night studies, in particular when behavioural and vehicle-based data is involved. Drowsiness is largely prominent at night \cite{RN128}, yet only 47\% of behavioural, vehicle, and hybrid papers included night scenarios in their study. Those that had night monitoring sometimes had systems that did not work in the daytime \cite{RN126}. Drowsiness can occur at any time and has often been shown to be present between 14:00 and 16:00 as well as at night \cite{RN10}. 

\subsection{Simulation versus On Road}
Within the literature, an absence of real and on-road studies was markedly present. Of the studies included in this survey, only 15 out of 126 had on-road data, often lacking drowsy participants. Earlier studies had more on-road data (11 out of the 15 prior to 2018), perhaps due to the lack of technology available for simulations. The on-road studies often had few participants and minimal drowsy participants (if any), where behavioural studies often asked subjects to “act” drowsy \cite{RN129, RN130}. Differences were reported between simulated and on-road studies, such as the effect of duration and perception of risk, furthering the need for on-road studies to be conducted \cite{RN10, RN131, RN132}.

\subsection{Loss of Data}
The loss of data can occur during recording as sensors can detach, weather can interfere with vehicle-based recordings or lighting variances can hinder behavioural recordings. When using an adaptable hybrid model or a model with multiple drowsiness feature sources, data loss does not have to mean ceased drowsiness monitoring. One study explored this, with a system that continued to work when a source signal was lost and drowsiness monitoring could continue \cite{RN133}. Data loss can occur at any time during monitoring, hence the system should continue to work if this is to occur.

\subsection{Sleep Disorders}
Sleep disorders have been reported to increase the chance of a driver being drowsy as they often affect the quality of sleep \cite{RN76}. Sleep disorders can also affect the results of drowsiness studies and hence most studies do not allow participants with sleep disorders to take part. However, this area still needs to be explored. For example, 33\% of all accidents in Australia have been related to sleep problems \cite{RN134}, and excluding these participants from studies is thus extremely problematic. 

One study explored the effects of obstructive sleep apnoea (OSAS), various sleep disorders (VSD) and “normals” in relation to the sleep-wake transition \cite{RN89}. The study collected EEG data, respiration, EMG and heart rate data, where the identified differences between subject types are summarised as follows:
\begin{itemize}
     \item Respiration frequency was different in the VSD group; however, the variability did not change across the three. 
     \item EMG results appeared to be the same across the groups.
     \item ECG R-R Intervals (RRI) had a correlation with drowsiness for the regular group and VSD group but not for the OSAS group.
     \item The very low frequency power component of RRI was consistent across all three groups.
     \item The low frequency power had a correlation with drowsiness for the normal and VSD group but not the OSAS group.
\end{itemize}

This study demonstrated that differences are seen between subject types and this should be explored when developing devices for on road use. Almost all other drowsiness studies omit these subjects from trials; however, driver drowsiness products cannot be considered effective until sleep disorder integration and monitoring is included. 

\subsection{Public Acceptance}
For behavioural monitoring in particular, public acceptance is a challenge for driver drowsiness technologies due to concerns regarding data storage and privacy protection. As camera monitoring is required, concerns of what else the data may be used for may hinder public acceptance of the product. Ways to counter this can include driver deidentification as described in \cite{RN140}. This can include extracting only relevant components required to determine if the driver is drowsy, such as eye and mouth segments. The rest of the image data can then be blurred, blackened or the components completely extracted from the image to make it harder to identify drivers.

\section{Conclusion and future directions}
Progressing driver drowsiness research requires current challenges to be addressed in order to develop reliable models for commercial use. We discussed how each of these modalities may be very much affected by noise and data loss/contact loss, ambient conditions, subject diversity etc. It is hoped that multi-modal data may alleviate this. Current research is promising in terms of accuracy and foundations; however, further research should be conducted to validate approaches for on road use. Adaption of models should also be considered, with the objective of earlier or pre-emptive driver drowsiness detection, before the driver is put at risk.  

A large number of challenges were presented in the literature, many of which relate to the data collection for driver drowsiness detection. Hence, for studies that are able to collect their own data, we emphasise that authors should be consistent and address the issues raised insofar as possible, with particular focus on the following:
\begin{itemize}
     \item Collecting on-road data where possible, with appropriate ethical considerations.
     \item Recruiting a diverse range and larger number of participants, with generalisability in mind.
     \item Labelling of drowsiness should be reasoned and if KSS is used, an adequate time frame between KSS samples should be maintained.
     \item Protocols for data collection needs to be thorough, with both day and night studies conducted for vehicle-based and behavioural-based studies in particular.
\end{itemize}
Furthermore, analysis of collected data needs to be addressed, with appropriate model validation. Proper window sizes for feature collection should be used in order to detect drowsiness in a timely manner. Prediction rather than detection should be used, in order to notify the driver in time, before an accident or adverse event occurs. Further research can include how far in advance drowsiness can be predicted and what methods are best suited for this task.

Hybrid models have shown to be more accurate, flexible and use the benefits of each individual method. They can allow for data loss and overcome the downfalls of intrusiveness, as a variety of sources can be used to collect data. However, they come at an increased complexity of development and cost. Further research should be undertaken for these approaches, bearing in mind the complexity they carry. 

Going forward, it is important for studies of drowsiness detection to also require subjects with sleep disorders to be included in studies, so they too can be monitored effectively on road. Once these challenges are addressed, accurate monitoring of drowsiness and hopefully more commercial implementation can be achieved.



\ifCLASSOPTIONcaptionsoff
  \newpage
\fi

\bibliographystyle{IEEEtran}

\bibliography{IEEEabrv, Main}

\begin{thebibliography}{100}
\providecommand{\url}[1]{#1}
\csname url@samestyle\endcsname
\providecommand{\newblock}{\relax}
\providecommand{\bibinfo}[2]{#2}
\providecommand{\BIBentrySTDinterwordspacing}{\spaceskip=0pt\relax}
\providecommand{\BIBentryALTinterwordstretchfactor}{4}
\providecommand{\BIBentryALTinterwordspacing}{\spaceskip=\fontdimen2\font plus
\BIBentryALTinterwordstretchfactor\fontdimen3\font minus
  \fontdimen4\font\relax}
\providecommand{\BIBforeignlanguage}[2]{{%
\expandafter\ifx\csname l@#1\endcsname\relax
\typeout{** WARNING: IEEEtran.bst: No hyphenation pattern has been}%
\typeout{** loaded for the language `#1'. Using the pattern for}%
\typeout{** the default language instead.}%
\else
\language=\csname l@#1\endcsname
\fi
#2}}
\providecommand{\BIBdecl}{\relax}
\BIBdecl

\bibitem{RN290}
\BIBentryALTinterwordspacing
``Road traffic injuries,'' World Health Organisation, 2021. [Online].
  Available:
  \url{https://www.who.int/news-room/fact-sheets/detail/road-traffic-injuries}
\BIBentrySTDinterwordspacing

\bibitem{RN5}
\BIBentryALTinterwordspacing
G.~Li and W.-Y. Chung, ``Detection of driver drowsiness using wavelet analysis
  of heart rate variability and a support vector machine classifier,''
  \emph{Sensors}, vol.~13, no.~12, pp. 16\,494--16\,511, 2013. [Online].
  Available: \url{https://www.mdpi.com/1424-8220/13/12/16494}
\BIBentrySTDinterwordspacing

\bibitem{RN6}
\BIBentryALTinterwordspacing
A.~M. Williamson and A.~M. Feyer, ``Moderate sleep deprivation produces
  impairments in cognitive and motor performance equivalent to legally
  prescribed levels of alcohol intoxication,'' \emph{Occupational and
  environmental medicine}, vol.~57, no.~10, pp. 649--655, 2000. [Online].
  Available: \url{https://pubmed.ncbi.nlm.nih.gov/10984335
  https://www.ncbi.nlm.nih.gov/pmc/articles/PMC1739867/}
\BIBentrySTDinterwordspacing

\bibitem{RN1}
\BIBentryALTinterwordspacing
M.~D. Mulhall, T.~L. Sletten, M.~Magee, J.~E. Stone, S.~Ganesan, A.~Collins,
  C.~Anderson, S.~W. Lockley, M.~E. Howard, and S.~M.~W. Rajaratnam,
  ``Sleepiness and driving events in shift workers: the impact of circadian and
  homeostatic factors,'' \emph{Sleep}, vol.~42, no.~6, 2019. [Online].
  Available: \url{https://doi.org/10.1093/sleep/zsz074}
\BIBentrySTDinterwordspacing

\bibitem{RN2}
\BIBentryALTinterwordspacing
CARRS-Q, ``Sleepiness and fatigue,'' 2018. [Online]. Available:
  \url{https://research.qut.edu.au/carrsq/wp-content/uploads/sites/296/2020/12/Sleepiness-and-fatigue.pdf}
\BIBentrySTDinterwordspacing

\bibitem{RN3}
\BIBentryALTinterwordspacing
2021. [Online]. Available:
  \url{https://www.rsc.wa.gov.au/RSC/media/Documents/Road%20Data/Statistics/WA-Road-Fatality-Progress-2020.pdf}
\BIBentrySTDinterwordspacing

\bibitem{RN4}
\BIBentryALTinterwordspacing
2020. [Online]. Available:
  \url{https://www.parliament.vic.gov.au/images/stories/committees/SCEI/Inquiry_into_the_Increase_in_Victorias_Road_Toll_/Submissions/S53_-_RACV_Redacted.pdf}
\BIBentrySTDinterwordspacing

\bibitem{RN135}
\BIBentryALTinterwordspacing
``Drivers are falling asleep behind the wheel,'' 2015. [Online]. Available:
  \url{https://www.nsc.org/road-safety/safety-topics/fatigued-driving}
\BIBentrySTDinterwordspacing

\bibitem{RN289}
\BIBentryALTinterwordspacing
``Drowsy driving,'' National Sleep Foundation. [Online]. Available:
  \url{https://drowsydriving.org/about/facts-and-stats/}
\BIBentrySTDinterwordspacing

\bibitem{RN10}
\BIBentryALTinterwordspacing
A.~Sahayadhas, K.~Sundaraj, and M.~Murugappan, ``Detecting driver drowsiness
  based on sensors: A review,'' \emph{Sensors}, vol.~12, no.~12, pp.
  16\,937--16\,953, 2012. [Online]. Available:
  \url{https://www.mdpi.com/1424-8220/12/12/16937}
\BIBentrySTDinterwordspacing

\bibitem{RN279}
S.~Kaplan, M.~A. Guvensan, A.~G. Yavuz, and Y.~Karalurt, ``Driver behavior
  analysis for safe driving: A survey,'' \emph{IEEE Transactions on Intelligent
  Transportation Systems}, vol.~16, no.~6, pp. 3017--3032, 2015.

\bibitem{RN36}
M.~Doudou, A.~Bouabdallah, and V.~Berge-Cherfaoui, ``Driver drowsiness
  measurement technologies: Current research, market solutions, and
  challenges,'' \emph{International Journal of Intelligent Transportation
  Systems Research}, pp. 1--23, 2019.

\bibitem{RN11}
M.~Ramzan, H.~U. Khan, S.~M. Awan, A.~Ismail, M.~Ilyas, and A.~Mahmood, ``A
  survey on state-of-the-art drowsiness detection techniques,'' \emph{IEEE
  Access}, vol.~7, pp. 61\,904--61\,919, 2019.

\bibitem{RN7}
\BIBentryALTinterwordspacing
X.~Hu and G.~Lodewijks, ``Detecting fatigue in car drivers and aircraft pilots
  by using non-invasive measures: The value of differentiation of sleepiness
  and mental fatigue,'' \emph{Journal of Safety Research}, vol.~72, pp.
  173--187, 2020. [Online]. Available:
  \url{https://www.sciencedirect.com/science/article/pii/S0022437519306735}
\BIBentrySTDinterwordspacing

\bibitem{RN29}
A.~Němcová, V.~Svozilová, K.~Bucsuházy, R.~Smíšek, M.~Mézl, B.~Hesko,
  M.~Belák, M.~Bilík, P.~Maxera, and M.~Seitl, ``Multimodal features for
  detection of driver stress and fatigue,'' \emph{IEEE Transactions on
  Intelligent Transportation Systems}, 2020.

\bibitem{RN22}
J.~C. Stutts, J.~W. Wilkins, and B.~V. Vaughn, ``Why do people have drowsy
  driving crashes,'' \emph{Input from drivers who just did}, vol. 202, no. 638,
  p. 5944, 1999.

\bibitem{RN23}
\BIBentryALTinterwordspacing
T.~ÅKERSTEDT, B.~PETERS, A.~ANUND, and G.~KECKLUND, ``Impaired alertness and
  performance driving home from the night shift: a driving simulator study,''
  \emph{Journal of Sleep Research}, vol.~14, no.~1, pp. 17--20, 2005. [Online].
  Available:
  \url{https://onlinelibrary.wiley.com/doi/abs/10.1111/j.1365-2869.2004.00437.x}
\BIBentrySTDinterwordspacing

\bibitem{RN24}
C.~M. D.~W. Roth~T, Roehrs~TA, ``Daytime sleepiness and alertness,''
  \emph{Principles and practice of sleep medicine}, pp. 14--23, 1994.

\bibitem{RN25}
M.~Ngxande, ``Correcting inter-sectional accuracy differences in drowsiness
  detection systems using generative adversarial networks (gans),'' Thesis,
  2020.

\bibitem{RN291}
\BIBentryALTinterwordspacing
``Fatigue,'' Better Health Channel. [Online]. Available:
  \url{https://www.betterhealth.vic.gov.au/health/conditionsandtreatments/fatigue}
\BIBentrySTDinterwordspacing

\bibitem{RN26}
J.~M. Lyznicki, T.~C. Doege, R.~M. Davis, and M.~A. Williams, ``Sleepiness,
  driving, and motor vehicle crashes,'' \emph{Jama}, vol. 279, no.~23, pp.
  1908--1913, 1998.

\bibitem{RN27}
\BIBentryALTinterwordspacing
K.~Cherney, ``How long does caffeine stay in your system?'' vol. 2021, no. 28
  May, 2018. [Online]. Available:
  \url{https://www.healthline.com/health/how-long-does-caffeine-last}
\BIBentrySTDinterwordspacing

\bibitem{RN28}
\BIBentryALTinterwordspacing
E.~De~Valck and R.~Cluydts, ``Slow-release caffeine as a countermeasure to
  driver sleepiness induced by partial sleep deprivation,'' \emph{Journal of
  Sleep Research}, vol.~10, no.~3, pp. 203--209, 2001. [Online]. Available:
  \url{https://onlinelibrary.wiley.com/doi/abs/10.1046/j.1365-2869.2001.00260.x}
\BIBentrySTDinterwordspacing

\bibitem{RN8}
A.~Chowdhury, R.~Shankaran, M.~Kavakli, and M.~M. Haque, ``Sensor applications
  and physiological features in drivers’ drowsiness detection: A review,''
  \emph{IEEE Sensors Journal}, vol.~18, no.~8, pp. 3055--3067, 2018.

\bibitem{RN9}
\BIBentryALTinterwordspacing
T.~Kundinger, N.~Sofra, and A.~Riener, ``Assessment of the potential of
  wrist-worn wearable sensors for driver drowsiness detection,'' \emph{Sensors
  (Basel, Switzerland)}, vol.~20, no.~4, p. 1029, 2020. [Online]. Available:
  \url{https://pubmed.ncbi.nlm.nih.gov/32075030
  https://www.ncbi.nlm.nih.gov/pmc/articles/PMC7070962/}
\BIBentrySTDinterwordspacing

\bibitem{RN12}
M.~Ngxande, J.~Tapamo, and M.~Burke, ``Driver drowsiness detection using
  behavioral measures and machine learning techniques: A review of state-of-art
  techniques,'' in \emph{2017 Pattern Recognition Association of South Africa
  and Robotics and Mechatronics (PRASA-RobMech)}, Conference Proceedings, pp.
  156--161.

\bibitem{RN140}
S.~Martin, A.~Tawari, and M.~M. Trivedi, ``Toward privacy-protecting safety
  systems for naturalistic driving videos,'' \emph{IEEE Transactions on
  Intelligent Transportation Systems}, vol.~15, no.~4, pp. 1811--1822, 2014.

\bibitem{RN13}
\BIBentryALTinterwordspacing
C.~C. Liu, S.~G. Hosking, and M.~G. Lenné, ``Predicting driver drowsiness
  using vehicle measures: Recent insights and future challenges,''
  \emph{Journal of Safety Research}, vol.~40, no.~4, pp. 239--245, 2009.
  [Online]. Available:
  \url{https://www.sciencedirect.com/science/article/pii/S0022437509000668}
\BIBentrySTDinterwordspacing

\bibitem{RN14}
\BIBentryALTinterwordspacing
P.~M. Forsman, B.~J. Vila, R.~A. Short, C.~G. Mott, and H.~P.~A. Van~Dongen,
  ``Efficient driver drowsiness detection at moderate levels of drowsiness,''
  \emph{Accident Analysis \& Prevention}, vol.~50, pp. 341--350, 2012.
  [Online]. Available:
  \url{https://www.sciencedirect.com/science/article/pii/S0001457512001571}
\BIBentrySTDinterwordspacing

\bibitem{RN15}
\BIBentryALTinterwordspacing
C.~N. Watling, M.~Mahmudul~Hasan, and G.~S. Larue, ``Sensitivity and
  specificity of the driver sleepiness detection methods using physiological
  signals: A systematic review,'' \emph{Accident Analysis \& Prevention}, vol.
  150, p. 105900, 2021. [Online]. Available:
  \url{https://www.sciencedirect.com/science/article/pii/S0001457520317206}
\BIBentrySTDinterwordspacing

\bibitem{RN16}
C.~Zhang, H.~Wang, and R.~Fu, ``Automated detection of driver fatigue based on
  entropy and complexity measures,'' \emph{IEEE Transactions on Intelligent
  Transportation Systems}, vol.~15, no.~1, pp. 168--177, 2014.

\bibitem{RN17}
S.~Kim, B.~Choi, T.~Cho, Y.~Lee, H.~Koo, and D.~Kim, ``Development of a
  classification model for driver's drowsiness and waking status using heart
  rate variability and respiratory features,'' \emph{Journal of the Ergonomics
  Society of Korea}, vol.~35, pp. 371--381, 2016.

\bibitem{RN18}
T.~Hwang, M.~Kim, S.~Hong, and K.~S. Park, ``Driver drowsiness detection using
  the in-ear eeg,'' in \emph{2016 38th Annual International Conference of the
  IEEE Engineering in Medicine and Biology Society (EMBC)}, Conference
  Proceedings, pp. 4646--4649.

\bibitem{RN19}
H.~Xue-Qin, W.~Zheng, and B.~Lu, ``Driving fatigue detection with fusion of eeg
  and forehead eog,'' in \emph{2016 International Joint Conference on Neural
  Networks (IJCNN)}, Conference Proceedings, pp. 897--904.

\bibitem{RN20}
\BIBentryALTinterwordspacing
M.~Awais, N.~Badruddin, and M.~Drieberg, ``A hybrid approach to detect driver
  drowsiness utilizing physiological signals to improve system performance and
  wearability,'' \emph{Sensors}, vol.~17, no.~9, p. 1991, 2017. [Online].
  Available: \url{https://www.mdpi.com/1424-8220/17/9/1991}
\BIBentrySTDinterwordspacing

\bibitem{RN21}
\BIBentryALTinterwordspacing
J.~Gielen and J.-M. Aerts, ``Feature extraction and evaluation for driver
  drowsiness detection based on thermoregulation,'' \emph{Applied Sciences},
  vol.~9, no.~17, p. 3555, 2019. [Online]. Available:
  \url{https://www.mdpi.com/2076-3417/9/17/3555}
\BIBentrySTDinterwordspacing

\bibitem{RN32}
\BIBentryALTinterwordspacing
``Driver vigilance telemetric control system - vigiton,'' 2013. [Online].
  Available: \url{http://www.neurocom.ru/en2/product/vigiton.html}
\BIBentrySTDinterwordspacing

\bibitem{RN33}
\BIBentryALTinterwordspacing
``Stopsleep,'' 2017. [Online]. Available: \url{https://www.stopsleep.com.au/}
\BIBentrySTDinterwordspacing

\bibitem{RN34}
\BIBentryALTinterwordspacing
``Steer: Wearable device that will not let you fall asleep,'' 2017. [Online].
  Available:
  \url{https://www.kickstarter.com/projects/creativemode/steer-you-will-never-fall-asleep-while-driving}
\BIBentrySTDinterwordspacing

\bibitem{RN35}
\BIBentryALTinterwordspacing
``Life by smartcap,'' 2016. [Online]. Available:
  \url{http://www.smartcaptech.com/}
\BIBentrySTDinterwordspacing

\bibitem{RN37}
N.~Edenborough, R.~Hammoud, A.~Harbach, A.~Ingold, B.~Kisacanin, P.~Malawey,
  T.~Newman, G.~Scharenbroch, S.~Skiver, and M.~Smith, ``Driver state monitor
  from delphi,'' in \emph{2005 IEEE Computer Society Conference on Computer
  Vision and Pattern Recognition (CVPR'05)}, vol.~2.\hskip 1em plus 0.5em minus
  0.4em\relax IEEE, Conference Proceedings, pp. 1206--1207.

\bibitem{RN38}
\BIBentryALTinterwordspacing
``Eagle light.'' [Online]. Available:
  \url{https://www.optalert.com/explore-products/scientifically-validated-glasses-mining/}
\BIBentrySTDinterwordspacing

\bibitem{RN39}
\BIBentryALTinterwordspacing
``Driver monitoring system (dms),'' 2016. [Online]. Available:
  \url{https://smarteye.se/automotive-solutions/}
\BIBentrySTDinterwordspacing

\bibitem{RN40}
\BIBentryALTinterwordspacing
``Eyealert distracted driving and fatigue warning systems.'' [Online].
  Available: \url{http://eyealert.com/index.html}
\BIBentrySTDinterwordspacing

\bibitem{RN41}
\BIBentryALTinterwordspacing
``Eyetracker warns against momentary driver drowsiness,'' 2010. [Online].
  Available:
  \url{https://www.fraunhofer.de/en/press/research-news/2010/10/eye-tracker-driver-drowsiness.html}
\BIBentrySTDinterwordspacing

\bibitem{RN42}
\BIBentryALTinterwordspacing
``Infrared led detects drivers in microsleep,'' vol. 2021, no. 21 May, 2010.
  [Online]. Available:
  \url{http://w1.siemens.com.cn/news_en/frontier_technology_en/1748.aspx}
\BIBentrySTDinterwordspacing

\bibitem{RN43}
\BIBentryALTinterwordspacing
``Hxgn mineprotect operator alertness system light vehicle,'' 2019. [Online].
  Available:
  \url{https://hexagonmining.com/solutions/safety-portfolio/hxgn-mineprotect-operator-alertness-system}
\BIBentrySTDinterwordspacing

\bibitem{RN44}
\BIBentryALTinterwordspacing
``Guardian.'' [Online]. Available:
  \url{https://www.seeingmachines.com/guardian/guardian/}
\BIBentrySTDinterwordspacing

\bibitem{RN45}
\BIBentryALTinterwordspacing
``Saab tech: Keep drivers awake and focused,'' vol. 2021, no. 24 May, 2014.
  [Online]. Available:
  \url{https://www.saabplanet.com/saab-tech-keep-drivers-awake-and-focused/}
\BIBentrySTDinterwordspacing

\bibitem{RN46}
\BIBentryALTinterwordspacing
``Intercore's driver alertness detection system™ now available,'' 2014.
  [Online]. Available:
  \url{https://www.newswire.com/news/intercores-driver-alertness-detection-system-now-available}
\BIBentrySTDinterwordspacing

\bibitem{RN47}
\BIBentryALTinterwordspacing
``Drivealert+.'' [Online]. Available:
  \url{https://driverisk.com.au/drivealert/}
\BIBentrySTDinterwordspacing

\bibitem{RN48}
\BIBentryALTinterwordspacing
``Driver fatigue monitoring system driving status detection.'' [Online].
  Available:
  \url{https://stonkam.com/products/Driver-Fatigue-Monitoring-System-AD-A11.html}
\BIBentrySTDinterwordspacing

\bibitem{RN49}
\BIBentryALTinterwordspacing
``Ts driver fatigue monitor dfm2.'' [Online]. Available:
  \url{http://www.transportsupport.co.uk/product/ts-driver-fatigue-monitor-dfm2/#.YKdBR6gzaUk}
\BIBentrySTDinterwordspacing

\bibitem{RN50}
\BIBentryALTinterwordspacing
``Fas100-user-guide-may2007,'' 2007. [Online]. Available:
  \url{http://service.alan-electronics.de/Archiv/Fahrer-Assistenz-Systeme/FAS%20100/FAS100-user-guide.pdf}
\BIBentrySTDinterwordspacing

\bibitem{RN51}
\BIBentryALTinterwordspacing
``Drowsy driver detection systems sense when you need a break,'' vol. 2021, no.
  24 May, 2016. [Online]. Available:
  \url{https://www.cars.com/articles/drowsy-driver-detection-systems-sense-when-you-need-a-break-1420684409199/}
\BIBentrySTDinterwordspacing

\bibitem{RN52}
\BIBentryALTinterwordspacing
``Driver alert control (dac),'' 2018. [Online]. Available:
  \url{https://www.volvocars.com/en-th/support/manuals/v60/2016w17/driver-support/driver-alert-system/driver-alert-control-dac}
\BIBentrySTDinterwordspacing

\bibitem{RN53}
\BIBentryALTinterwordspacing
``Cognex vision helps vehicle “see” the road in 2007 darpa urban
  challenge,'' 2007. [Online]. Available:
  \url{https://www.cognex.com/company/press-releases/2007/cognex-vision-helps-vehicle-see-the-road-in-2007-darpa-urban-challenge}
\BIBentrySTDinterwordspacing

\bibitem{RN54}
\BIBentryALTinterwordspacing
``Lexus safety system+.'' [Online]. Available:
  \url{https://drivers.lexus.com/lexus-drivers-theme/pdf/LSS+%20Quick%20Guide%20Link.pdf}
\BIBentrySTDinterwordspacing

\bibitem{RN55}
\BIBentryALTinterwordspacing
``Subaru eyesight.'' [Online]. Available:
  \url{https://www.subaru.com/engineering/eyesight.html}
\BIBentrySTDinterwordspacing

\bibitem{RN56}
\BIBentryALTinterwordspacing
``Astid® summary of operation,'' fmi, Report. [Online]. Available:
  \url{https://fmiapplications.com/astid.html}
\BIBentrySTDinterwordspacing

\bibitem{RN57}
\BIBentryALTinterwordspacing
``Iteris announces introduction of industry-first lane departure warning data
  collection product "safety direct i\& trade" for the heavy truck market,''
  vol. 2021, no. 24 May, 2008. [Online]. Available:
  \url{https://www.iteris.com/news/iteris-announces-introduction-industry-first-lane-departure-warning-data-collection-product}
\BIBentrySTDinterwordspacing

\bibitem{RN58}
\BIBentryALTinterwordspacing
``Driver fatigue detection.'' [Online]. Available:
  \url{https://www.milescontinental.co.nz/news/features/driver-fatigue-detection/}
\BIBentrySTDinterwordspacing

\bibitem{RN30}
\BIBentryALTinterwordspacing
``Toyota launches ls and mirai equipped with "advanced drive" that enables
  drivers and cars to drive together in japan,'' vol. 2021, no. 24 May, 2021.
  [Online]. Available:
  \url{https://global.toyota/en/newsroom/corporate/35063150.html}
\BIBentrySTDinterwordspacing

\bibitem{RN31}
\BIBentryALTinterwordspacing
``Interior monitoring systems.'' [Online]. Available:
  \url{https://www.bosch-mobility-solutions.com/en/solutions/interior/interior-monitoring-systems/}
\BIBentrySTDinterwordspacing

\bibitem{RN59}
L.~Oliveira, J.~S. Cardoso, A.~Lourenço, and C.~Ahlström, ``Driver drowsiness
  detection: a comparison between intrusive and non-intrusive signal
  acquisition methods,'' in \emph{2018 7th European Workshop on Visual
  Information Processing (EUVIP)}, Conference Proceedings, pp. 1--6.

\bibitem{RN60}
J.~Krajewski, D.~Sommer, U.~Trutschel, D.~Edwards, and M.~Golz, ``Steering
  wheel behavior based estimation of fatigue,'' 2009.

\bibitem{RN61}
A.~Lemkaddem, R.~Delgado-Gonzalo, E.~Türetken, S.~Dasen, V.~Moser, C.~Gressum,
  J.~Solà, D.~Ferrario, and C.~Verjus, ``Multi-modal driver drowsiness
  detection: A feasibility study,'' in \emph{2018 IEEE EMBS International
  Conference on Biomedical \& Health Informatics (BHI)}, Conference
  Proceedings, pp. 9--12.

\bibitem{RN62}
\BIBentryALTinterwordspacing
S.~Arefnezhad, S.~Samiee, A.~Eichberger, and A.~Nahvi, ``Driver drowsiness
  detection based on steering wheel data applying adaptive neuro-fuzzy feature
  selection,'' \emph{Sensors}, vol.~19, no.~4, p. 943, 2019. [Online].
  Available: \url{https://www.mdpi.com/1424-8220/19/4/943}
\BIBentrySTDinterwordspacing

\bibitem{RN63}
P.~Ebrahim, ``Driver drowsiness monitoring using eye movement features derived
  from electrooculography,'' 2016.

\bibitem{RN64}
J.~Ma, J.~Zhang, Z.~Gong, and Y.~Du, ``Study on fatigue driving detection model
  based on steering operation features and eye movement features,'' in
  \emph{2018 IEEE 4th International Conference on Control Science and Systems
  Engineering (ICCSSE)}, Conference Proceedings, pp. 472--475.

\bibitem{RN65}
\BIBentryALTinterwordspacing
M.~Chai, s.-w. Li, w.-c. Sun, m.-z. Guo, and m.-y. Huang, ``Drowsiness
  monitoring based on steering wheel status,'' \emph{Transportation Research
  Part D: Transport and Environment}, vol.~66, pp. 95--103, 2019. [Online].
  Available:
  \url{https://www.sciencedirect.com/science/article/pii/S1361920917306582}
\BIBentrySTDinterwordspacing

\bibitem{RN66}
\BIBentryALTinterwordspacing
X.~Zhang, X.~Wang, X.~Yang, C.~Xu, X.~Zhu, and J.~Wei, ``Driver drowsiness
  detection using mixed-effect ordered logit model considering time cumulative
  effect,'' \emph{Analytic Methods in Accident Research}, vol.~26, p. 100114,
  2020. [Online]. Available:
  \url{https://www.sciencedirect.com/science/article/pii/S221366572030004X}
\BIBentrySTDinterwordspacing

\bibitem{RN67}
M.~H. Baccour, F.~Driewer, T.~Schäck, and E.~Kasneci, ``Camera-based driver
  drowsiness state classification using logistic regression models,'' in
  \emph{2020 IEEE International Conference on Systems, Man, and Cybernetics
  (SMC)}, Conference Proceedings, pp. 1--8.

\bibitem{RN68}
B.~Thorslund, \emph{Electrooculogram analysis and development of a system for
  defining stages of drowsiness}.\hskip 1em plus 0.5em minus 0.4em\relax
  Statens väg-och transportforskningsinstitut, 2004.

\bibitem{RN69}
\BIBentryALTinterwordspacing
R.~Li, Y.~V. Chen, and L.~Zhang, ``A method for fatigue detection based on
  driver's steering wheel grip,'' \emph{International Journal of Industrial
  Ergonomics}, vol.~82, p. 103083, 2021. [Online]. Available:
  \url{https://www.sciencedirect.com/science/article/pii/S0169814121000019}
\BIBentrySTDinterwordspacing

\bibitem{RN70}
M.~W. Johns, ``A new method for measuring daytime sleepiness: the epworth
  sleepiness scale,'' \emph{sleep}, vol.~14, no.~6, pp. 540--545, 1991.

\bibitem{RN71}
\BIBentryALTinterwordspacing
L.~L. Di~Stasi, R.~Renner, A.~Catena, J.~J. Cañas, B.~M. Velichkovsky, and
  S.~Pannasch, ``Towards a driver fatigue test based on the saccadic main
  sequence: A partial validation by subjective report data,''
  \emph{Transportation Research Part C: Emerging Technologies}, vol.~21, no.~1,
  pp. 122--133, 2012. [Online]. Available:
  \url{https://www.sciencedirect.com/science/article/pii/S0968090X1100101X}
\BIBentrySTDinterwordspacing

\bibitem{RN72}
M.~Cella and T.~Chalder, ``Measuring fatigue in clinical and community
  settings,'' \emph{Journal of psychosomatic research}, vol.~69, no.~1, pp.
  17--22, 2010.

\bibitem{RN73}
T.~Chalder, G.~Berelowitz, T.~Pawlikowska, L.~Watts, S.~Wessely, D.~Wright, and
  E.~Wallace, ``Development of a fatigue scale,'' \emph{Journal of
  psychosomatic research}, vol.~37, no.~2, pp. 147--153, 1993.

\bibitem{RN74}
R.~J. Hanowski, M.~Blanco, A.~Nakata, J.~S. Hickman, W.~A. Schaudt, M.~Fumero,
  R.~L. Olson, J.~Jermeland, M.~Greening, and G.~Holbrook, ``The drowsy driver
  warning system field operational test: Data collection methods,'' 2008.

\bibitem{RN75}
\BIBentryALTinterwordspacing
A.~D. McDonald, J.~D. Lee, C.~Schwarz, and T.~L. Brown, ``A contextual and
  temporal algorithm for driver drowsiness detection,'' \emph{Accident Analysis
  \& Prevention}, vol. 113, pp. 25--37, 2018. [Online]. Available:
  \url{https://www.sciencedirect.com/science/article/pii/S0001457518300058}
\BIBentrySTDinterwordspacing

\bibitem{RN76}
S.~E.~H. Kiashari, A.~Nahvi, A.~Homayounfard, and H.~Bakhoda, ``Monitoring the
  variation in driver respiration rate from wakefulness to drowsiness: a
  non-intrusive method for drowsiness detection using thermal imaging,''
  \emph{Journal of Sleep Sciences}, vol.~3, no. 1-2, pp. 1--9, 2018.

\bibitem{RN77}
M.~Mahmoodi and A.~Nahvi, ``Driver drowsiness detection based on classification
  of surface electromyography features in a driving simulator,''
  \emph{Proceedings of the Institution of Mechanical Engineers, Part H: Journal
  of Engineering in Medicine}, vol. 233, no.~4, pp. 395--406, 2019.

\bibitem{RN78}
A.~Eskandarian and A.~Mortazavi, ``Evaluation of a smart algorithm for
  commercial vehicle driver drowsiness detection,'' in \emph{2007 IEEE
  Intelligent Vehicles Symposium}, Conference Proceedings, pp. 553--559.

\bibitem{RN79}
M.~W. Johns, A.~Tucker, R.~Chapman, K.~Crowley, and N.~Michael, ``Monitoring
  eye and eyelid movements by infrared reflectance oculography to measure
  drowsiness in drivers,'' \emph{Somnologie-Schlafforschung und Schlafmedizin},
  vol.~11, no.~4, pp. 234--242, 2007.

\bibitem{RN80}
\BIBentryALTinterwordspacing
``The science.'' [Online]. Available:
  \url{https://www.optalert.com/why-optalert/science/}
\BIBentrySTDinterwordspacing

\bibitem{RN81}
M.~V. Ramesh, A.~K. Nair, and A.~T. Kunnathu, ``Real-time automated multiplexed
  sensor system for driver drowsiness detection,'' in \emph{2011 7th
  International Conference on Wireless Communications, Networking and Mobile
  Computing}, Conference Proceedings, pp. 1--4.

\bibitem{RN82}
Y.~Lin, H.~Leng, G.~Yang, and H.~Cai, ``An intelligent noninvasive sensor for
  driver pulse wave measurement,'' \emph{IEEE Sensors Journal}, vol.~7, no.~5,
  pp. 790--799, 2007.

\bibitem{RN83}
L.~Yong~Gyu, K.~Ko~Keun, and P.~Suk, ``Ecg measurement on a chair without
  conductive contact,'' \emph{IEEE Transactions on Biomedical Engineering},
  vol.~53, no.~5, pp. 956--959, 2006.

\bibitem{RN156}
Z.~Gao, X.~Wang, Y.~Yang, C.~Mu, Q.~Cai, W.~Dang, and S.~Zuo, ``Eeg-based
  spatio–temporal convolutional neural network for driver fatigue
  evaluation,'' \emph{IEEE Transactions on Neural Networks and Learning
  Systems}, vol.~30, no.~9, pp. 2755--2763, 2019.

\bibitem{RN157}
Z.-K. Gao, Y.-L. Li, Y.-X. Yang, and C.~Ma, ``A recurrence network-based
  convolutional neural network for fatigue driving detection from eeg,''
  \emph{Chaos: An Interdisciplinary Journal of Nonlinear Science}, vol.~29,
  no.~11, p. 113126, 2019.

\bibitem{RN159}
C.~T. Lin, C.~H. Chuang, Y.~C. Hung, C.~N. Fang, D.~Wu, and Y.~K. Wang, ``A
  driving performance forecasting system based on brain dynamic state analysis
  using 4-d convolutional neural networks,'' \emph{IEEE Transactions on
  Cybernetics}, pp. 1--9, 2020.

\bibitem{RN155}
\BIBentryALTinterwordspacing
K.~T. Chui, M.~D. Lytras, and R.~W. Liu, ``A generic design of driver
  drowsiness and stress recognition using moga optimized deep mkl-svm,''
  \emph{Sensors}, vol.~20, no.~5, p. 1474, 2020. [Online]. Available:
  \url{https://www.mdpi.com/1424-8220/20/5/1474}
\BIBentrySTDinterwordspacing

\bibitem{RN161}
\BIBentryALTinterwordspacing
S.~Murugan, J.~Selvaraj, and A.~Sahayadhas, ``Detection and analysis: driver
  state with electrocardiogram (ecg),'' \emph{Physical and Engineering Sciences
  in Medicine}, vol.~43, no.~2, pp. 525--537, 2020. [Online]. Available:
  \url{https://doi.org/10.1007/s13246-020-00853-8}
\BIBentrySTDinterwordspacing

\bibitem{RN154}
T.~Chang, ``Detection of exercise fatigue using neural network with grey
  relational analysis from hrv signal,'' in \emph{2019 IEEE International
  Conference on Computation, Communication and Engineering (ICCCE)}, 2019,
  Conference Proceedings, pp. 87--90.

\bibitem{RN92}
L.~Boon-Leng, L.~Dae-Seok, and L.~Boon-Giin, ``Mobile-based wearable-type of
  driver fatigue detection by gsr and emg,'' in \emph{TENCON 2015 - 2015 IEEE
  Region 10 Conference}, Conference Proceedings, pp. 1--4.

\bibitem{RN127}
A.~A. Hayawi and J.~Waleed, ``Driver's drowsiness monitoring and alarming
  auto-system based on eog signals,'' in \emph{2019 2nd International
  Conference on Engineering Technology and its Applications (IICETA)},
  Conference Proceedings, pp. 214--218.

\bibitem{RN160}
J.-X. Ma, L.-C. Shi, and B.~Lu, ``An eog-based vigilance estimation method
  applied for driver fatigue detection,'' 2015, Conference Proceedings.

\bibitem{RN150}
\BIBentryALTinterwordspacing
M.~Sidikova, R.~Martinek, A.~Kawala-Sterniuk, M.~Ladrova, R.~Jaros, L.~Danys,
  and P.~Simonik, ``Vital sign monitoring in car seats based on
  electrocardiography, ballistocardiography and seismocardiography: A review,''
  \emph{Sensors}, vol.~20, no.~19, 2020. [Online]. Available:
  \url{https://www.mdpi.com/1424-8220/20/19/5699}
\BIBentrySTDinterwordspacing

\bibitem{RN158}
\BIBentryALTinterwordspacing
H.~Lee, J.~Lee, and M.~Shin, ``Using wearable ecg/ppg sensors for driver
  drowsiness detection based on distinguishable pattern of recurrence plots,''
  \emph{Electronics}, vol.~8, no.~2, p. 192, 2019. [Online]. Available:
  \url{https://www.mdpi.com/2079-9292/8/2/192}
\BIBentrySTDinterwordspacing

\bibitem{RN162}
F.~Trenta, S.~Conoci, F.~Rundo, and S.~Battiato, ``Advanced motion-tracking
  system with multi-layers deep learning framework for innovative car-driver
  drowsiness monitoring,'' in \emph{2019 14th IEEE International Conference on
  Automatic Face \& Gesture Recognition (FG 2019)}, Conference Proceedings, pp.
  1--5.

\bibitem{RN84}
L.~Chin-Teng, W.~Ruei-Cheng, L.~Sheng-Fu, C.~Wen-Hung, C.~Yu-Jie, and
  J.~Tzyy-Ping, ``Eeg-based drowsiness estimation for safety driving using
  independent component analysis,'' \emph{IEEE Transactions on Circuits and
  Systems I: Regular Papers}, vol.~52, no.~12, pp. 2726--2738, 2005.

\bibitem{RN85}
G.~Li, B.~Lee, and W.~Chung, ``Smartwatch-based wearable eeg system for driver
  drowsiness detection,'' \emph{IEEE Sensors Journal}, vol.~15, no.~12, pp.
  7169--7180, 2015.

\bibitem{RN86}
A.~Mehreen, S.~M. Anwar, M.~Haseeb, M.~Majid, and M.~O. Ullah, ``A hybrid
  scheme for drowsiness detection using wearable sensors,'' \emph{IEEE Sensors
  Journal}, vol.~19, no.~13, pp. 5119--5126, 2019.

\bibitem{RN95}
S.~Koh, B.~R. Cho, J.~Lee, S.~Kwon, S.~Lee, J.~B. Lim, S.~B. Lee, and H.~Kweon,
  ``Driver drowsiness detection via ppg biosignals by using multimodal head
  support,'' in \emph{2017 4th International Conference on Control, Decision
  and Information Technologies (CoDIT)}, Conference Proceedings, pp.
  0383--0388.

\bibitem{RN96}
H.~Rahim, A.~Dalimi, and H.~Jaafar, ``Detecting drowsy driver using pulse
  sensor,'' \emph{Jurnal Teknologi}, vol.~73, 2015.

\bibitem{RN151}
R.~Favilla, V.~C. Zuccalà, and G.~Coppini, ``Heart rate and heart rate
  variability from single-channel video and ica integration of multiple
  signals,'' \emph{IEEE Journal of Biomedical and Health Informatics}, vol.~23,
  no.~6, pp. 2398--2408, 2019.

\bibitem{RN152}
F.~Zhao, M.~Li, Y.~Qian, and J.~Z. Tsien, ``Remote measurements of heart and
  respiration rates for telemedicine,'' \emph{PloS one}, vol.~8, no.~10, p.
  e71384, 2013.

\bibitem{RN87}
U.~Svensson, \emph{Blink behaviour based drowsiness detection: method
  development and validation}.\hskip 1em plus 0.5em minus 0.4em\relax Statens
  väg-och transportforskningsinstitut, 2004.

\bibitem{RN88}
\BIBentryALTinterwordspacing
D.~SHIN, H.~SAKAI, and Y.~UCHIYAMA, ``Slow eye movement detection can prevent
  sleep-related accidents effectively in a simulated driving task,''
  \emph{Journal of Sleep Research}, vol.~20, no.~3, pp. 416--424, 2011.
  [Online]. Available:
  \url{https://onlinelibrary.wiley.com/doi/abs/10.1111/j.1365-2869.2010.00891.x}
\BIBentrySTDinterwordspacing

\bibitem{RN89}
\BIBentryALTinterwordspacing
Z.~Shinar, S.~Akselrod, Y.~Dagan, and A.~Baharav, ``Autonomic changes during
  wake–sleep transition: A heart rate variability based approach,''
  \emph{Autonomic Neuroscience}, vol. 130, no.~1, pp. 17--27, 2006. [Online].
  Available:
  \url{https://www.sciencedirect.com/science/article/pii/S1566070206001135}
\BIBentrySTDinterwordspacing

\bibitem{RN90}
\BIBentryALTinterwordspacing
I.~Hostens and H.~Ramon, ``Assessment of muscle fatigue in low level monotonous
  task performance during car driving,'' \emph{Journal of Electromyography and
  Kinesiology}, vol.~15, no.~3, pp. 266--274, 2005. [Online]. Available:
  \url{https://www.sciencedirect.com/science/article/pii/S1050641104000823}
\BIBentrySTDinterwordspacing

\bibitem{RN91}
\BIBentryALTinterwordspacing
V.~Balasubramanian and K.~Adalarasu, ``Emg-based analysis of change in muscle
  activity during simulated driving,'' \emph{Journal of Bodywork and Movement
  Therapies}, vol.~11, no.~2, pp. 151--158, 2007. [Online]. Available:
  \url{https://www.sciencedirect.com/science/article/pii/S1360859207000034}
\BIBentrySTDinterwordspacing

\bibitem{RN93}
\BIBentryALTinterwordspacing
M.~Misbhauddin, A.~AlMutlaq, A.~Almithn, N.~Alshukr, and M.~Aleesa, ``Real-time
  driver drowsiness detection using wearable technology,'' in \emph{Proceedings
  of the 4th International Conference on Smart City Applications}.\hskip 1em
  plus 0.5em minus 0.4em\relax Association for Computing Machinery, Conference
  Proceedings, p. Article 94. [Online]. Available:
  \url{https://doi.org/10.1145/3368756.3369081}
\BIBentrySTDinterwordspacing

\bibitem{RN94}
\BIBentryALTinterwordspacing
S.~Bando, K.~Oiwa, and A.~Nozawa, ``Evaluation of dynamics of forehead skin
  temperature under induced drowsiness,'' \emph{IEEJ Transactions on Electrical
  and Electronic Engineering}, vol.~12, no.~S1, pp. S104--S109, 2017. [Online].
  Available: \url{https://onlinelibrary.wiley.com/doi/abs/10.1002/tee.22423}
\BIBentrySTDinterwordspacing

\bibitem{RN97}
G.~S. Maximous and H.~A. Bastawrous, ``Driver drowsiness detection based on
  humantenna effect for automotive safety systems,'' in \emph{2020 IEEE 9th
  Global Conference on Consumer Electronics (GCCE)}, Conference Proceedings,
  pp. 391--392.

\bibitem{RN98}
A.~Mittal, K.~Kumar, S.~Dhamija, and M.~Kaur, ``Head movement-based driver
  drowsiness detection: A review of state-of-art techniques,'' in \emph{2016
  IEEE International Conference on Engineering and Technology (ICETECH)},
  Conference Proceedings, pp. 903--908.

\bibitem{RN99}
A.~Joshi, S.~Kyal, S.~Banerjee, and T.~Mishra, ``In-the-wild drowsiness
  detection from facial expressions,'' in \emph{2020 IEEE Intelligent Vehicles
  Symposium (IV)}, Conference Proceedings, pp. 207--212.

\bibitem{RN100}
\BIBentryALTinterwordspacing
H.~Zhang, C.~Wu, Z.~Huang, X.~Yan, and T.~Z. Qiu, ``Sensitivity of lane
  position and steering angle measurements to driver fatigue,''
  \emph{Transportation Research Record}, vol. 2585, no.~1, pp. 67--76, 2016.
  [Online]. Available: \url{https://doi.org/10.3141/2585-08}
\BIBentrySTDinterwordspacing

\bibitem{RN101}
A.~Bamidele, K.~Kamardin, N.~Syazarin, S.~Mohd, I.~Shafi, A.~Azizan, N.~Aini,
  and H.~Mad, ``Non-intrusive driver drowsiness detection based on face and eye
  tracking,'' \emph{Int J. Adv. Comput. Sci. Appl}, vol.~10, pp. 549--569,
  2019.

\bibitem{RN102}
S.~Pritchett, E.~Zilberg, Z.~M. Xu, M.~Karrar, D.~Burton, and S.~Lal,
  ``Comparing accuracy of two algorithms for detecting driver drowsiness —
  single source (eeg) and hybrid (eeg and body movement),'' in \emph{7th
  International Conference on Broadband Communications and Biomedical
  Applications}, Conference Proceedings, pp. 179--184.

\bibitem{RN103}
J.~Gwak, M.~Shino, and A.~Hirao, ``Early detection of driver drowsiness
  utilizing machine learning based on physiological signals, behavioral
  measures, and driving performance,'' in \emph{2018 21st International
  Conference on Intelligent Transportation Systems (ITSC)}, Conference
  Proceedings, pp. 1794--1800.

\bibitem{RN104}
H.~De~Rosario, J.~S. Solaz, N.~Rodrıguez, and L.~M. Bergasa, ``Controlled
  inducement and measurement of drowsiness in a driving simulator,'' \emph{IET
  intelligent transport systems}, vol.~4, no.~4, pp. 280--288, 2010.

\bibitem{RN105}
\BIBentryALTinterwordspacing
J.~Gwak, A.~Hirao, and M.~Shino, ``An investigation of early detection of
  driver drowsiness using ensemble machine learning based on hybrid sensing,''
  \emph{Applied Sciences}, vol.~10, no.~8, p. 2890, 2020. [Online]. Available:
  \url{https://www.mdpi.com/2076-3417/10/8/2890}
\BIBentrySTDinterwordspacing

\bibitem{RN106}
\BIBentryALTinterwordspacing
C.~Jacobé~de Naurois, C.~Bourdin, A.~Stratulat, E.~Diaz, and J.-L. Vercher,
  ``Detection and prediction of driver drowsiness using artificial neural
  network models,'' \emph{Accident Analysis \& Prevention}, vol. 126, pp.
  95--104, 2019. [Online]. Available:
  \url{https://www.sciencedirect.com/science/article/pii/S0001457517304347}
\BIBentrySTDinterwordspacing

\bibitem{RN107}
\BIBentryALTinterwordspacing
C.~Craye, A.~Rashwan, M.~S. Kamel, and F.~Karray, ``A multi-modal driver
  fatigue and distraction assessment system,'' \emph{International Journal of
  Intelligent Transportation Systems Research}, vol.~14, no.~3, pp. 173--194,
  2016. [Online]. Available: \url{https://doi.org/10.1007/s13177-015-0112-9}
\BIBentrySTDinterwordspacing

\bibitem{RN108}
G.~D. Furman, A.~Baharav, C.~Cahan, and S.~Akselrod, ``Early detection of
  falling asleep at the wheel: A heart rate variability approach,'' in
  \emph{2008 Computers in Cardiology}, Conference Proceedings, pp. 1109--1112.

\bibitem{RN109}
Y.~Sun, X.~Yu, J.~Berilla, Z.~Liu, and G.~Wu, ``An in-vehicle physiological
  signal monitoring system for driver fatigue detection,'' Conference
  Proceedings.

\bibitem{RN110}
\BIBentryALTinterwordspacing
C.~Zhao, M.~Zhao, J.~Liu, and C.~Zheng, ``Electroencephalogram and
  electrocardiograph assessment of mental fatigue in a driving simulator,''
  \emph{Accident Analysis \& Prevention}, vol.~45, pp. 83--90, 2012. [Online].
  Available:
  \url{https://www.sciencedirect.com/science/article/pii/S0001457511003241}
\BIBentrySTDinterwordspacing

\bibitem{RN111}
P.~P. Caffier, U.~Erdmann, and P.~Ullsperger, ``Experimental evaluation of
  eye-blink parameters as a drowsiness measure,'' \emph{European journal of
  applied physiology}, vol.~89, no.~3, pp. 319--325, 2003.

\bibitem{RN112}
F.~Zhang, J.~Su, L.~Geng, and Z.~Xiao, ``Driver fatigue detection based on eye
  state recognition,'' in \emph{2017 International Conference on Machine Vision
  and Information Technology (CMVIT)}, Conference Proceedings, pp. 105--110.

\bibitem{RN113}
R.~Schleicher, N.~Galley, S.~Briest, and L.~Galley, ``Blinks and saccades as
  indicators of fatigue in sleepiness warnings: looking tired?''
  \emph{Ergonomics}, vol.~51, no.~7, pp. 982--1010, 2008.

\bibitem{RN114}
N.~S. Karuppusamy and B.~Kang, ``Multimodal system to detect driver fatigue
  using eeg, gyroscope, and image processing,'' \emph{IEEE Access}, vol.~8, pp.
  129\,645--129\,667, 2020.

\bibitem{RN115}
S.~Lawoyin, X.~Liu, D.~Fei, and O.~Bai, ``Detection methods for a low-cost
  accelerometer-based approach for driver drowsiness detection,'' in \emph{2014
  IEEE International Conference on Systems, Man, and Cybernetics (SMC)},
  Conference Proceedings, pp. 1636--1641.

\bibitem{RN116}
J.~Buolamwini and T.~Gebru, ``Gender shades: Intersectional accuracy
  disparities in commercial gender classification,'' in \emph{Conference on
  fairness, accountability and transparency}.\hskip 1em plus 0.5em minus
  0.4em\relax PMLR, Conference Proceedings, pp. 77--91.

\bibitem{RN117}
C.-H. Weng, Y.-H. Lai, and S.-H. Lai, ``Driver drowsiness detection via a
  hierarchical temporal deep belief network,'' in \emph{Asian Conference on
  Computer Vision}.\hskip 1em plus 0.5em minus 0.4em\relax Springer, Conference
  Proceedings, pp. 117--133.

\bibitem{RN164}
M.~Ngxande, J.-R. Tapamo, and M.~Burke, ``Bias remediation in driver drowsiness
  detection systems using generative adversarial networks,'' \emph{IEEE
  Access}, vol.~8, pp. 55\,592--55\,601, 2020.

\bibitem{RN118}
S.~Arefnezhad, A.~Eichberger, M.~Frühwirth, C.~Kaufmann, and M.~Moser,
  ``Driver drowsiness classification using data fusion of vehicle-based
  measures and ecg signals,'' in \emph{2020 IEEE International Conference on
  Systems, Man, and Cybernetics (SMC)}, Conference Proceedings, pp. 451--456.

\bibitem{RN119}
N.~Kuamr and N.~C. Barwar, ``Analysis of real time driver fatigue detection
  based on eye and yawning,'' Conference Proceedings.

\bibitem{RN120}
``The 2020 vision index,'' Optometry Australia, Report, 2020.

\bibitem{RN121}
\BIBentryALTinterwordspacing
S.~Jamshidi, R.~Azmi, M.~Sharghi, and M.~Soryani, ``Hierarchical deep neural
  networks to detect driver drowsiness,'' \emph{Multimedia Tools and
  Applications}, 2021. [Online]. Available:
  \url{https://doi.org/10.1007/s11042-021-10542-7}
\BIBentrySTDinterwordspacing

\bibitem{RN122}
R.~Jabbar, M.~Shinoy, M.~Kharbeche, K.~Al-Khalifa, M.~Krichen, and K.~Barkaoui,
  ``Driver drowsiness detection model using convolutional neural networks
  techniques for android application,'' in \emph{2020 IEEE International
  Conference on Informatics, IoT, and Enabling Technologies (ICIoT)},
  Conference Proceedings, pp. 237--242.

\bibitem{RN123}
\BIBentryALTinterwordspacing
R.~Jabbar, K.~Al-Khalifa, M.~Kharbeche, W.~Alhajyaseen, M.~Jafari, and
  S.~Jiang, ``Real-time driver drowsiness detection for android application
  using deep neural networks techniques,'' \emph{Procedia Computer Science},
  vol. 130, pp. 400--407, 2018. [Online]. Available:
  \url{https://www.sciencedirect.com/science/article/pii/S1877050918304137}
\BIBentrySTDinterwordspacing

\bibitem{RN124}
F.~Dornaika, J.~Reta, I.~Arganda-Carreras, and A.~Moujahid, ``Driver drowsiness
  detection in facial images,'' in \emph{2018 Eighth International Conference
  on Image Processing Theory, Tools and Applications (IPTA)}, Conference
  Proceedings, pp. 1--6.

\bibitem{RN125}
O.~Khunpisuth, T.~Chotchinasri, V.~Koschakosai, and N.~Hnoohom, ``Driver
  drowsiness detection using eye-closeness detection,'' in \emph{2016 12th
  International Conference on Signal-Image Technology \& Internet-Based Systems
  (SITIS)}.\hskip 1em plus 0.5em minus 0.4em\relax IEEE, Conference
  Proceedings, pp. 661--668.

\bibitem{RN126}
L.~M. Bergasa, J.~Nuevo, M.~A. Sotelo, R.~Barea, and M.~E. Lopez, ``Real-time
  system for monitoring driver vigilance,'' \emph{IEEE Transactions on
  Intelligent Transportation Systems}, vol.~7, no.~1, pp. 63--77, 2006.

\bibitem{RN128}
B.~Lee, B.~Lee, and W.~Chung, ``Standalone wearable driver drowsiness detection
  system in a smartwatch,'' \emph{IEEE Sensors Journal}, vol.~16, no.~13, pp.
  5444--5451, 2016.

\bibitem{RN129}
\BIBentryALTinterwordspacing
J.~Jo, S.~J. Lee, K.~R. Park, I.-J. Kim, and J.~Kim, ``Detecting driver
  drowsiness using feature-level fusion and user-specific classification,''
  \emph{Expert Systems with Applications}, vol.~41, no. 4, Part 1, pp.
  1139--1152, 2014. [Online]. Available:
  \url{https://www.sciencedirect.com/science/article/pii/S0957417413006106}
\BIBentrySTDinterwordspacing

\bibitem{RN130}
\BIBentryALTinterwordspacing
I.-H. Choi, C.-H. Jeong, and Y.-G. Kim, ``Tracking a driver’s face against
  extreme head poses and inference of drowsiness using a hidden markov model,''
  \emph{Applied Sciences}, vol.~6, no.~5, p. 137, 2016. [Online]. Available:
  \url{https://www.mdpi.com/2076-3417/6/5/137}
\BIBentrySTDinterwordspacing

\bibitem{RN131}
P.~Philip, P.~Sagaspe, N.~Moore, J.~Taillard, A.~Charles, C.~Guilleminault, and
  B.~Bioulac, ``Fatigue, sleep restriction and driving performance,''
  \emph{Accident Analysis \& Prevention}, vol.~37, no.~3, pp. 473--478, 2005.

\bibitem{RN132}
E.~Blana and J.~Golias, ``Differences between vehicle lateral displacement on
  the road and in a fixed-base simulator,'' \emph{Human factors}, vol.~44,
  no.~2, pp. 303--313, 2002.

\bibitem{RN133}
\BIBentryALTinterwordspacing
S.~Samiee, S.~Azadi, R.~Kazemi, A.~Nahvi, and A.~Eichberger, ``Data fusion to
  develop a driver drowsiness detection system with robustness to signal
  loss,'' \emph{Sensors}, vol.~14, no.~9, pp. 17\,832--17\,847, 2014. [Online].
  Available: \url{https://www.mdpi.com/1424-8220/14/9/17832}
\BIBentrySTDinterwordspacing

\bibitem{RN134}
S.~J. Motlagh, M.~Shabany, K.~S. Haghighi, A.~N. Nasrabadi, and S.~H.~E.
  Razavi, ``Relationship between sleep quality, obstructive sleep apnea and
  sleepiness during day with related factors in professional drivers,''
  \emph{Acta Medica Iranica}, pp. 690--695, 2017.

\end{thebibliography}

\begin{IEEEbiography}[{\includegraphics[width=1in,height=1.25in,clip,keepaspectratio]{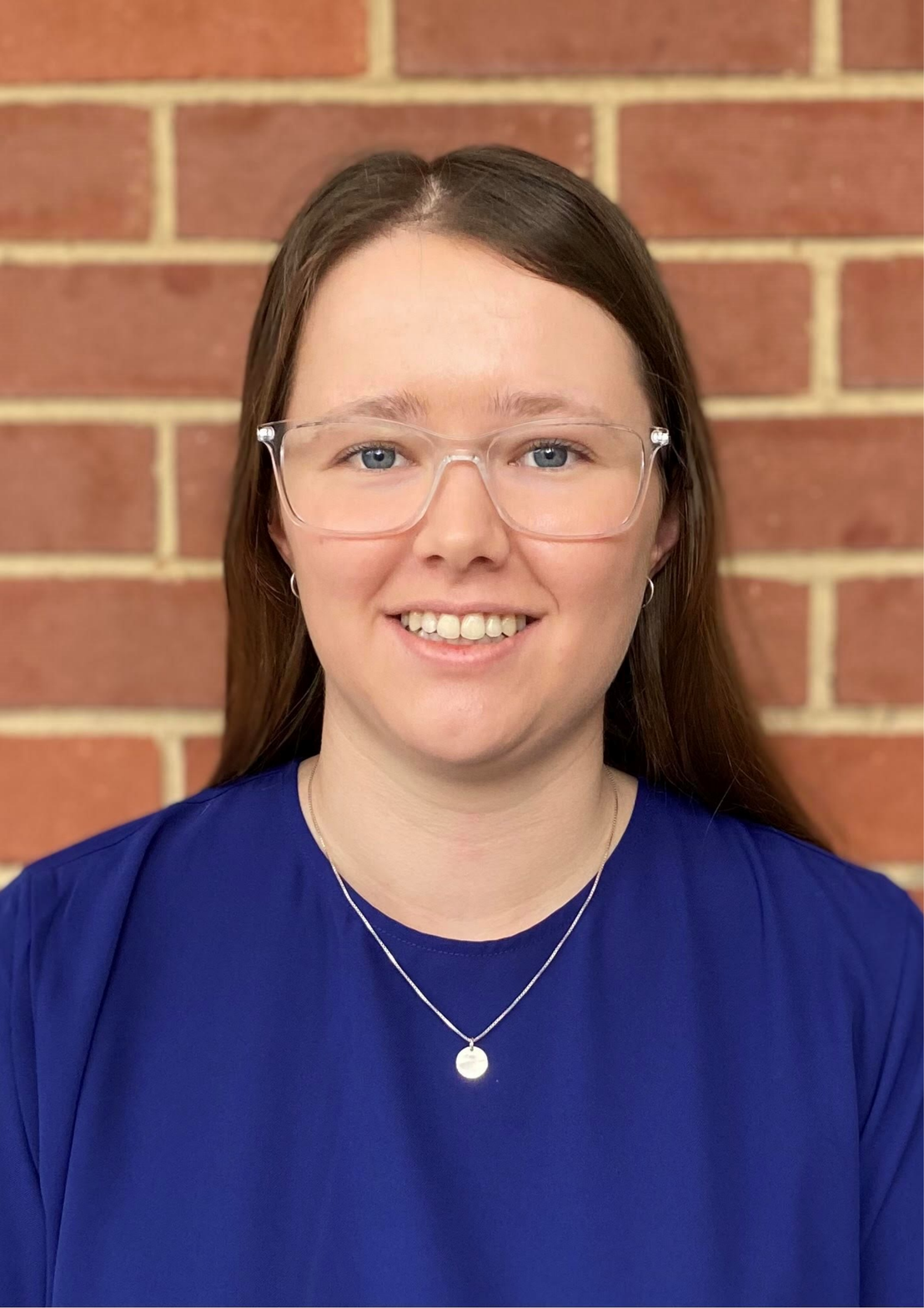}}]{EMMA PERKINS}

is a PhD Candidate in the department of electrical and computer systems engineering at Monash University, Melbourne, Australia. She received her B. Eng. in biomedical engineering from RMIT University, Melbourne, Australia in 2020. She is currently researching and developing a driver drowsiness detection system. Her interest areas include physiological monitoring, machine learning, biomedical devices and computer vision. 
\end{IEEEbiography}

\begin{IEEEbiography}[{\includegraphics[width=1in,height=1.25in,clip, trim=8cm 1cm 8cm 1cm, keepaspectratio]{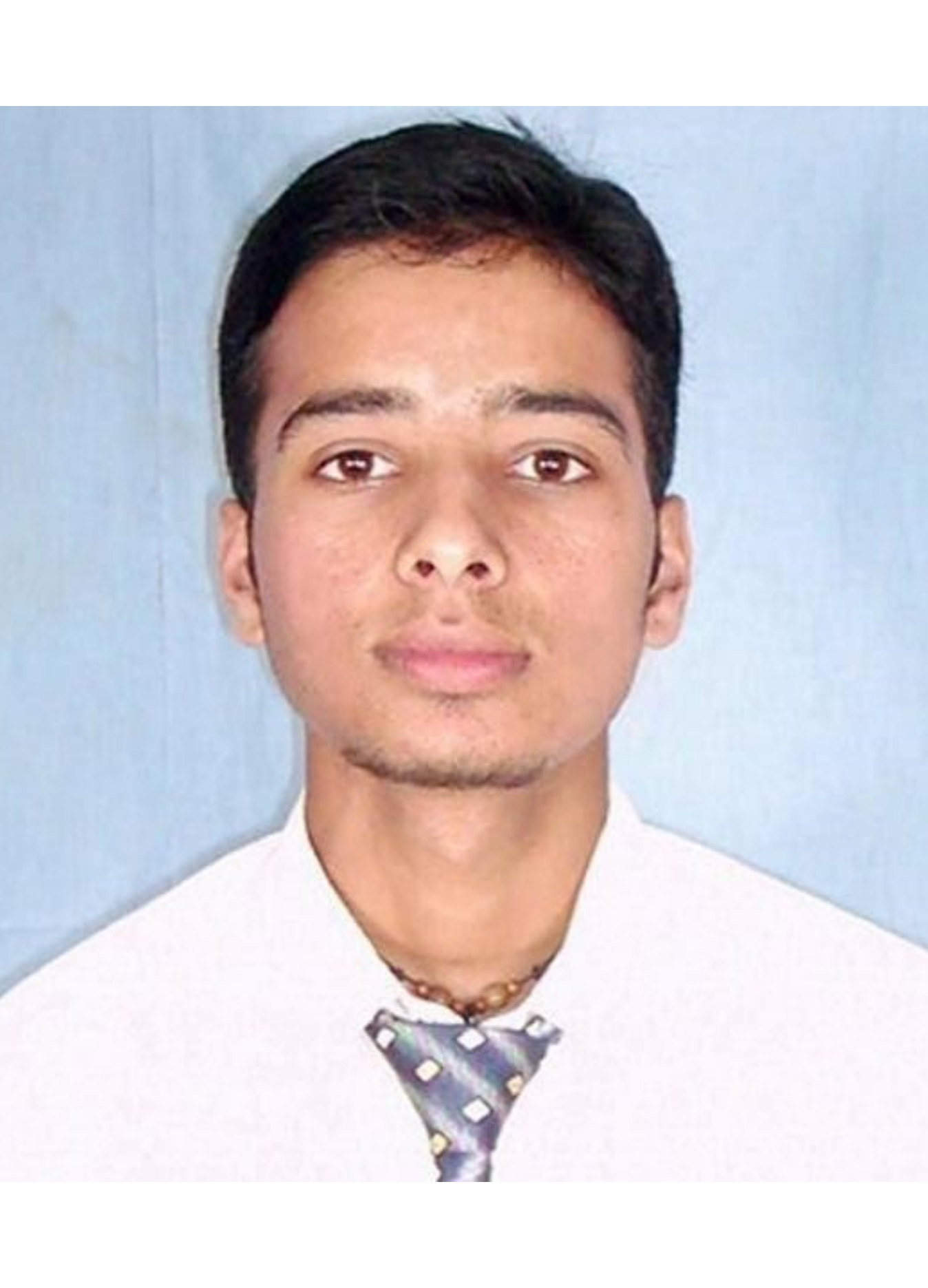}}]{CHIRANJIBI SITAULA}
 is a research fellow under the department of electrical and computer systems engineering, Monash University, Australia. He received a Ph.D. degree from Deakin University, Australia in 2021. He worked in industry and academia in Nepal for a number of years before joining Deakin University for a PhD degree. He has published research articles in top-tier conferences and journals in the field of deep learning and machine learning. His research interests include computer vision, signal processing, and machine learning. 
\end{IEEEbiography}

\begin{IEEEbiography}[{\includegraphics[width=1in,height=1.25in,clip,keepaspectratio]{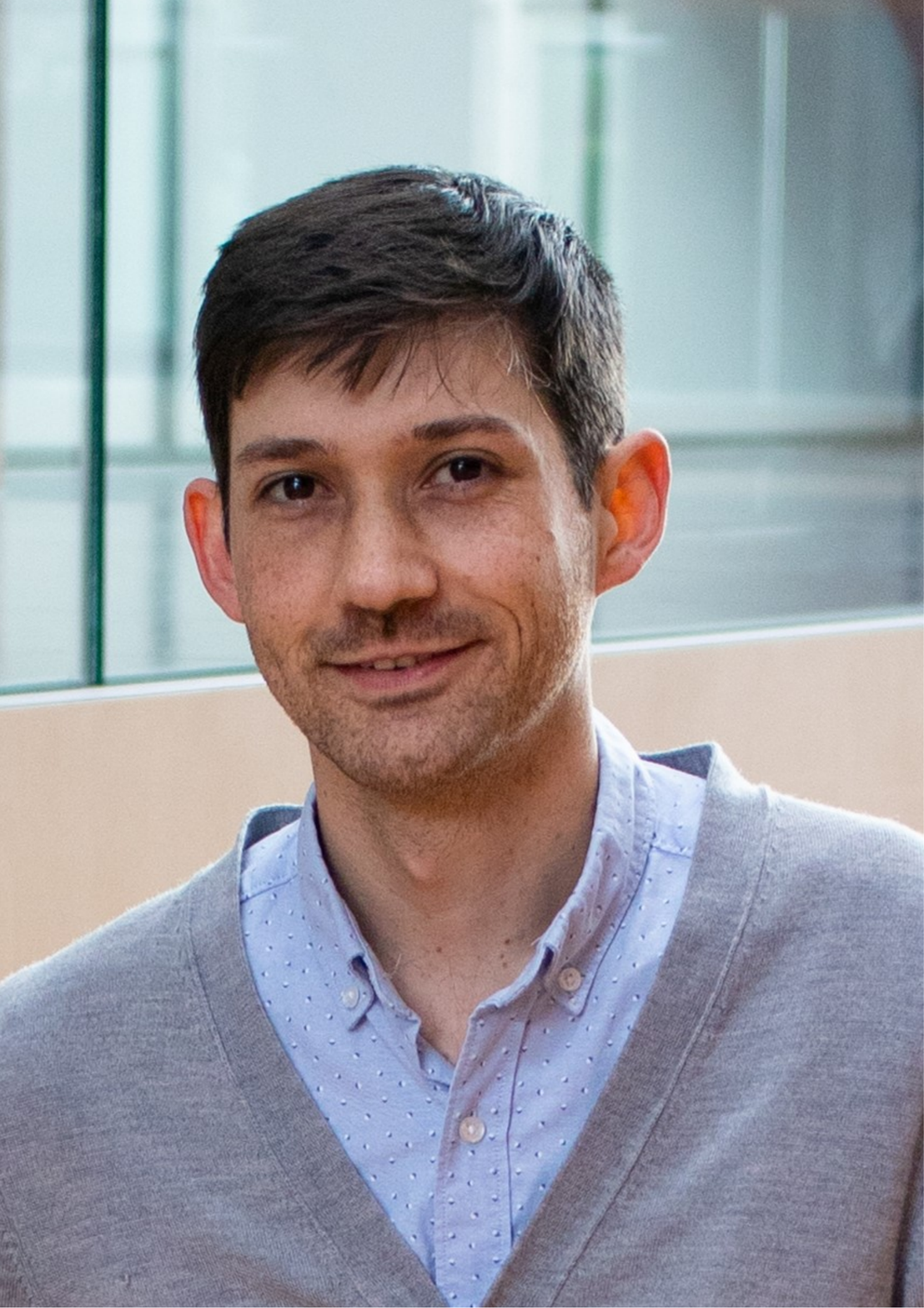}}]{MICHAEL BURKE} received a Ph.D in statistical signal processing from the University of Cambridge in 2016. He is  a lecturer specialising in computer vision, machine learning and control for autonomous robots, in the department of electrical and computer systems engineering, Monash University, Australia. He was a research associate working on robot learning at the University of Edinburgh, from 2018-2020. Before this, Michael led the Mobile Intelligent Autonomous Systems group at the Council for Scientific and Industrial Research (CSIR), South Africa. 
\end{IEEEbiography}

\begin{IEEEbiography}[{\includegraphics[width=1in,height=1.25in,clip,trim=5cm 10cm 6cm 7.5cm, keepaspectratio]{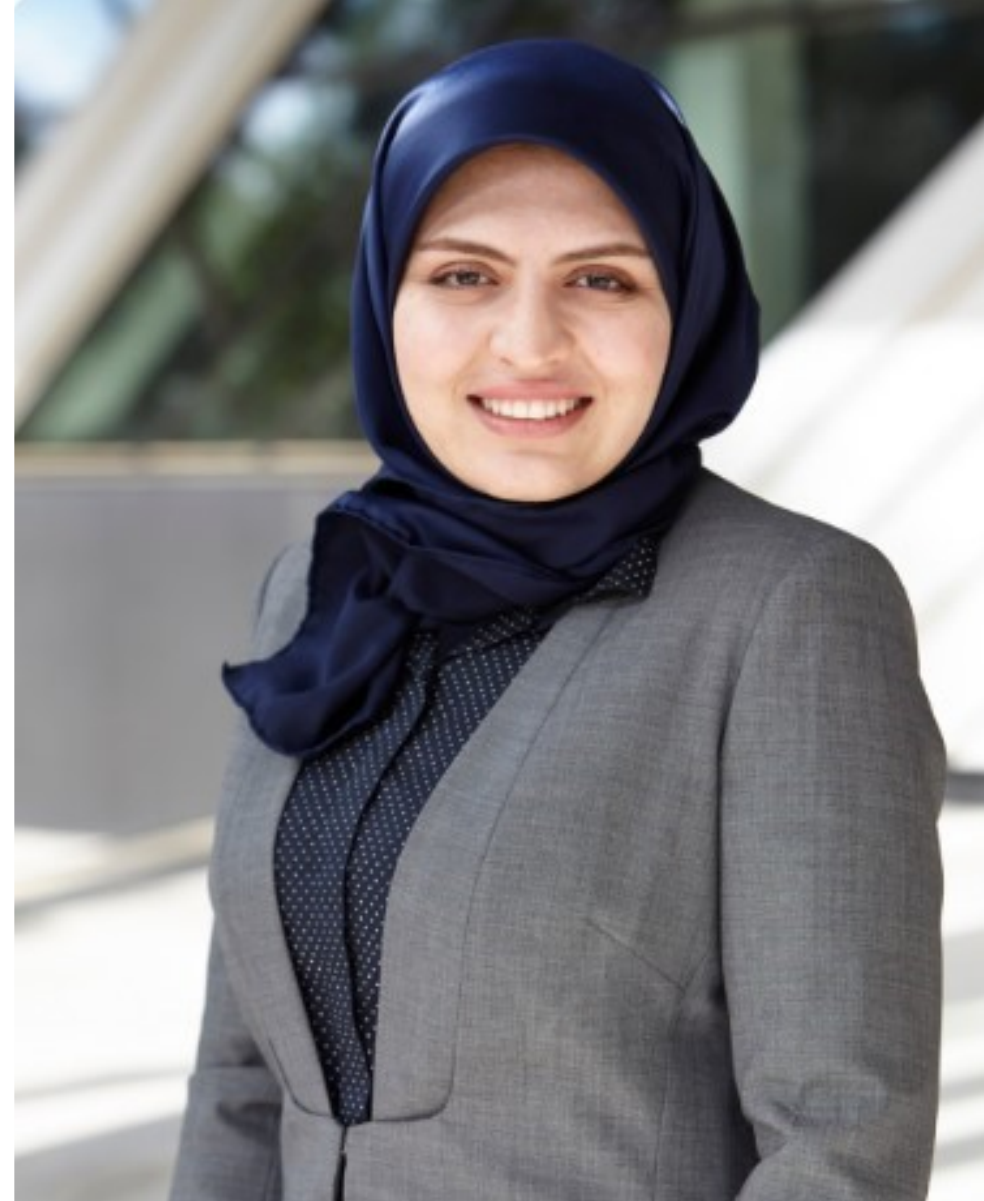}}]{FAEZEH MARZBANRAD}
received her PhD from University of Melbourne, Australia in 2016. She is currently a senior member of IEEE, lecturer and head of Biomedical Signal Processing Lab in the department of electrical and computer systems engineering, Monash University, Australia. Her research interests include biomedical signal processing, machine learning, affordable medical technologies and mobile-health. 
\end{IEEEbiography}

\end{document}